\documentclass[twocolumn]{aastex631}

\usepackage{graphicx}	
\usepackage{amsmath}	

\usepackage{amssymb}	

\newcommand{\Msun}{${\rm M}_{\odot}$}
\newcommand{\kms}{km\,s$^{-1}$}
\newcommand{\SiII}{Si~{\sc ii}}

\newcommand{\NaI}{Na~{\sc i}~D}

\newcommand{\OI}{O~{\sc i}}

\newcommand{\OIII}{O~{\sc iii}}

\newcommand{\FeVII}{Fe~{\sc vii}}

\newcommand{\HeI}{He~{\sc i}}
\newcommand{\HeII}{He~{\sc ii}}

\newcommand{\FeII}{Fe~{\sc ii}}
\newcommand{\NeIII}{Ne~{\sc iii}}

\newcommand{\CaII}{Ca~{\sc ii}}
\newcommand{\MgII}{Mg~{\sc ii}}

\newcommand{\Nifs}{$^{56}$Ni}

\shorttitle{SN 2014C}
\shortauthors{Zhai et al.}
\graphicspath{{./}{figures/}}

\begin{document}

\title{SN 2014C: a metamorphic supernova exploded in the intricate and hydrogen-rich surroundings}

\author{Qian Zhai}
\affiliation{Yunnan Observatories (YNAO), Chinese Academy of Sciences (CAS), Kunming, 650216, China}
\affiliation{Key Laboratory for the Structure and Evolution of Celestial Objects, CAS, Kunming, 650216, China}

\author[0000-0002-8296-2590]{Jujia Zhang}\thanks{E-mail:jujia@ynao.ac.cn}
\affiliation{Yunnan Observatories (YNAO), Chinese Academy of Sciences (CAS), Kunming, 650216, China}
\affiliation{Key Laboratory for the Structure and Evolution of Celestial Objects, CAS, Kunming, 650216, China}
\affiliation{International Centre of Supernovae, Yunnan Key Laboratory, Kunming 650216, China}

\author{Weili Lin}
\affiliation{Department of Astronomy, Xiamen University, Xiamen, Fujian 361005, China}
\affiliation{Physics Department and Tsinghua Center for Astrophysics (THCA), Tsinghua University, Beijing 100084, China}

\author{Paolo Mazzali}
\affiliation{Astrophysics Research Institute, Liverpool John Moores University, Liverpool Science Park, 146 Brownlow Hill, Liverpool L3 5RF, UK}
\affiliation{Max-Planck-Institut f\"ur Astrophysik, Karl-Schwarzschild Stra\ss{}e 1, 85748 Garching, Germany}

\author[0000-0001-8646-4858]{Elena Pian}
\affiliation{INAF, Astrophysics and Space Science Observatory, Via P. Gobetti 101, 40129 Bologna, Italy}

\author[0000-0002-3256-0016]{Stefano Benetti}
\affiliation{INAF-Osservatorio Astronomico di Padova, Vicolo dell'Osservatorio 5, I-35122 Padova, Italy}

\author[0000-0002-3697-2616]{Lina Tomasella}
\affiliation{INAF-Osservatorio Astronomico di Padova, Vicolo dell'Osservatorio 5, I-35122 Padova, Italy}

\author{Jialian Liu} 
\affiliation{Physics Department and Tsinghua Center for Astrophysics (THCA), Tsinghua University, Beijing 100084, China}

\author{Liping Li}
\affiliation{Yunnan Observatories (YNAO), Chinese Academy of Sciences (CAS), Kunming, 650216, China}
\affiliation{Key Laboratory for the Structure and Evolution of Celestial Objects, CAS, Kunming, 650216, China}
\affiliation{International Centre of Supernovae, Yunnan Key Laboratory, Kunming 650216, China}

\begin{abstract}
We present photometric and spectroscopic observations of supernova (SN) 2014C, primarily emphasizing the initial month after the explosion at approximately daily intervals. During this time, it was classified as a Type Ib SN exhibiting a notably higher peak luminosity ($L_{\rm peak}\approx4.3\times10^{42}\rm erg\,s^{-1}$), a faster rise to brightness ($t_{\rm rise} \approx 11.6$ d), and a more gradual dimming ($\Delta m_{15}^{V} \approx 0.48$ mag) compared to typical SNe Ib. Analysis of the velocity evolution over the first $\sim$ 20 days after the explosion supports the view that the absorption near 6200\AA\,is due to high-velocity H$\alpha$ in the outer layers of the ejecta, indicating the presence of a small amount of hydrogen in the envelope of progenitor before the explosion. Assuming the peak luminosity is entirely attributed to radioactive decay, we estimate that 0.14 \Msun\,of \Nifs\,was synthesized in the explosion. However, this amount of nickel could no longer maintain observed brightness approximately ten days after peak luminosity, suggesting additional energy sources beyond radioactive decay. This supplementary energy likely originates from interaction with the circumstellar medium (CSM). Consequently, the timing of the SN-CSM interaction in SN 2014C may occur much earlier than the emergence of IIn-like features during the nebular phase.
\end{abstract}

%
%


\section{Introduction} \label{sec:intro}

Stripped-envelope supernovae (SESNe) are the explosion of a massive star that has lost its hydrogen or helium envelopes either through stellar winds \citep{2014A&A...564A..30G} or via mass transfer to a companion star \citep{2010ApJ...725..940Y}. However, the processes by which a massive progenitor undergoes envelope stripping during the core collapse and the state of the circumstellar material (CSM) at the time of the certain SESNe explosion remain highly complex and not fully understood.

Recent observations of stellar explosions show that massive stars often experience complex mass loss before death. These include pre-explosion eruptions in H-rich stars, progenitors of ordinary Type IIn SNe \citep{2014ApJ...789..104O, 2013ApJ...779L...8F, 2013MNRAS.430.1801M,2013MNRAS.431.2599M,2014MNRAS.442.1166M,2013ApJ...767....1P, 2013ApJ...763L..27P,2014ARA&A..52..487S} as well as progenitors of Type IIp SNe that experience short-lived interactions with CSM \citep{2017A&A...603A..51D,2017NatPh..13..510Y,2024ApJ...970..189J,2023SciBu..68.2548Z,2024ApJ...970L..18Z}. Even H-poor progenitors show erratic mass-loss behavior preceding core-collapse \citep{2011ApJ...732...32F, 2007Natur.447..829P, 2008ApJ...674L..85I, 2014Natur.509..471G, 2015ApJ...807...35M}. This time-dependent mass loss differs from the steady loss assumed in current models. The cause of this highly variable mass loss is debated, and its role in the evolution of the progenitor remains unclear.

\begin{figure*}[ht!]
\plotone{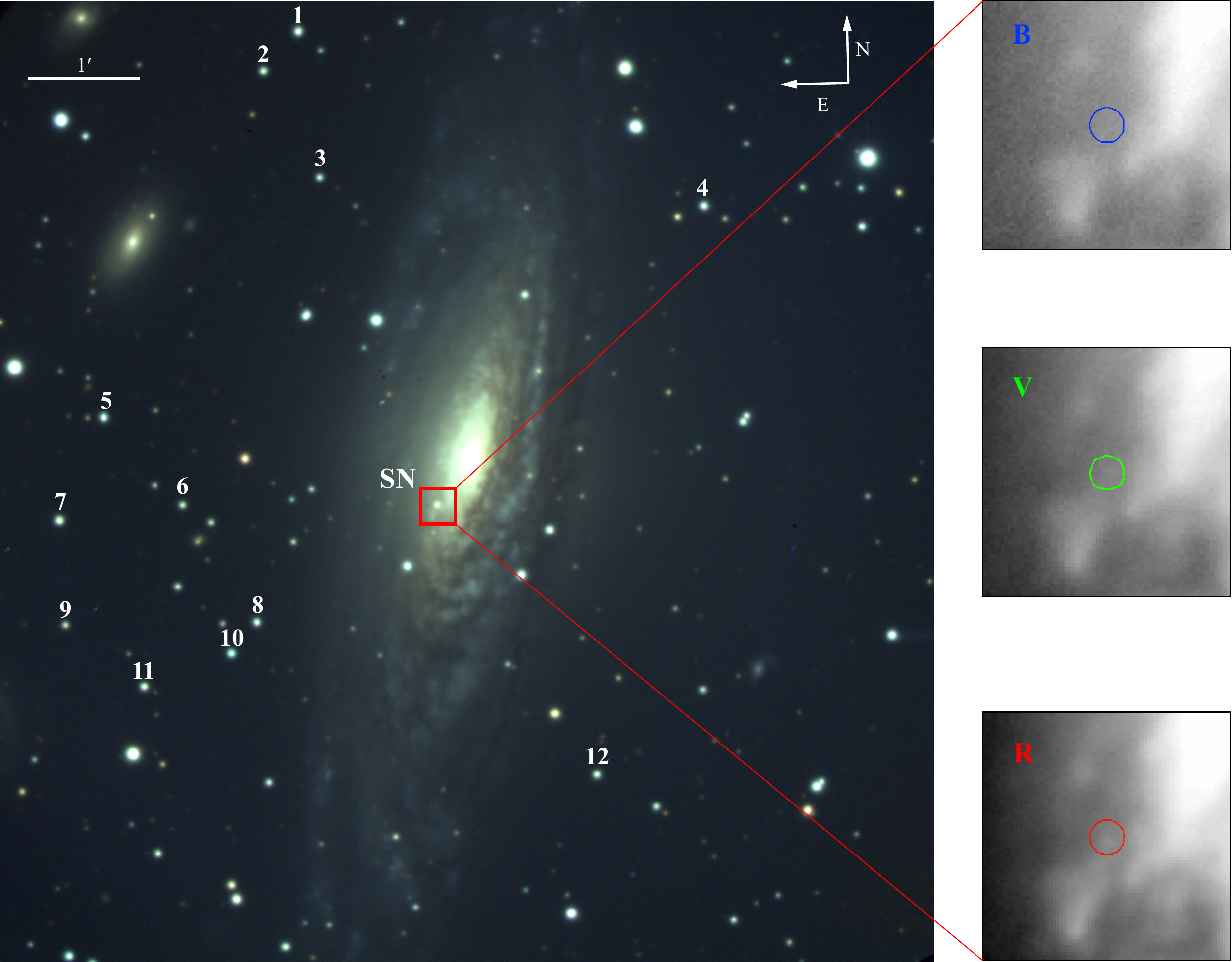}
\caption{Finder chart of SN\,2014C and its local reference stars, combined by $BVR$-bands images taken by the LJT and YFOSC in January 2014. The mean FWHM of the combined image is $\sim 1\arcsec.50$ under the scale of $\sim 0\arcsec.28$/pixel. The zoomed-in templates taken in November 2020 for the $BVR$-bands, which encompass the region containing SN 2014C, are presented on the right.  }
\label{<img>}
\end{figure*}

SN 2014C showed a remarkable metamorphosis from a hydrogen-poor (H-poor) Type Ib SN to an SN interacting with hydrogen-rich (H-rich) CSM, characterized by the intermediate broadening of H$\alpha$ emission line at about 100 days after the explosion \citep{2017ApJ...835..140M,2015ApJ...815..120M}. This shift arises from the interaction between SN and CSM, presenting a rare chance to study the mass loss of its progenitor leading up to the explosion. Some investigations on the later phase of SN 2014C have been published, particularly regarding the properties of sounding CSM derived by the interaction signal \citep{2017MNRAS.466.3648A, 2018MNRAS.475.1756B,2021MNRAS.502.1694B,2016ApJ...833..231T,2019ApJ...887...75T,2022ApJ...939..105B,2022ApJ...930...57T}. These observations suggest that SN 2014C exploded inside a low-density cavity, and the expanding shock encountered a dense H-rich shell within a year after the explosion. 

The transition from SNe Ib/c to SNe IIn, attributed to interactions with H-rich CSM, has been documented in multiple cases, including SN 2001em \citep{2006ApJ...641.1051C, 2020ApJ...902...55C}, SN 2004dk \citep{2018MNRAS.478.5050M}, SN 2019oys \citep{2020A&A...643A..79S}, and SN 2019yvr \citep{2022MNRAS.510.3701S, 2024MNRAS.529L..33F}, among others. These SNe exhibit varying timing and intensity of interactions, indicative of differences in the geometry and density of CSM.

However, the relationship between these changes and SN explosion parameters needs to be confirmed. This uncertainty primarily arises from our limited understanding of the explosion properties of these peculiar SNe. They usually attract attention later when significant interaction signals are detected, with insufficient early observations. For instance, SN 2001em and SN 2019oys have only one low-quality identification spectrum from their early phases.

SN 2014C posed a similar challenge. Shortly after the explosion, its position behind the Sun restricted observations, limiting researcher interest. Notably, only two early spectra have been published \citep{2015ApJ...815..120M,2022ApJ...930...57T}, which is insufficient for determining the evolutionary trend during the SN Ib phase. Furthermore, early photometric results \citep{2017ApJ...835..140M} may have overestimated the luminosity of this SN, particularly during its rise and decline, due to the absence of host galaxy template subtraction.

Here, we present spectra from the first month after the explosion, including nearly daily observations in the initial two weeks. Through template subtraction using pre- and post-explosion images, we have improved the photometry of this SN. Analysis of these observations enables us to investigate many explosion properties, which are crucial for understanding the mechanisms of this extraordinary event.

The manuscript is organized as follows: observations and data reductions are described in Section \ref{sect:obs}; Section \ref{lc&cc} presents the UV and optical light and color curves, while Section \ref{spectra} presents the spectral evolution. In Section \ref{bolo}, we calculate and analyze the bolometric light curve of SN 2014C. Finally, a summary follows in Section \ref{sum}.

\begin{figure*}[ht!]
\plotone{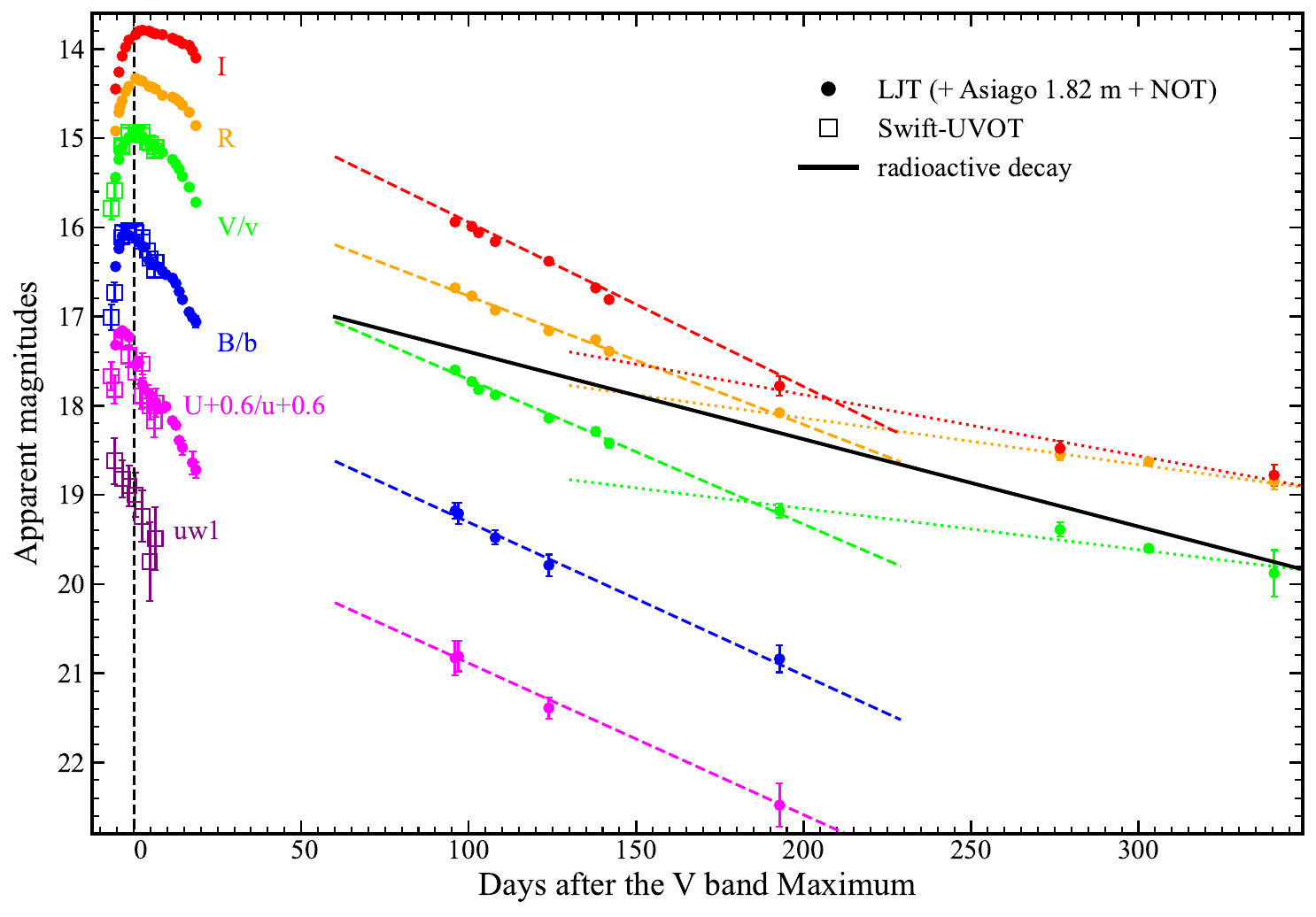}
\caption{$UBVRI$- bands and $uw1, u, b, v$- bands light curves of SN 2014C. The dashed- and dotted lines are the linear fit of the $UBVRI$- band light curve at the tail phase. The solid line is the radioactive decay rate (i.e., 0.98 mag / 100d) from $^{56}\rm Co$ to $^{56}\rm Fe$. }
\label{<14C_LC>}
\end{figure*}

\section{Observations and Data Reduction} \label{sect:obs}

SN 2014C was discovered in an unfiltered image of the spiral galaxy NGC 7331 \citep{2014ATel.5721....1Z} taken on Jan. 5.09 UT 2014 (UT is used throughout this paper) using the 0.76-m Katzman Automatic Imaging Telescope (KAIT) at Lick Observatory. It was also detected three days earlier on a pre-discovery image taken on Jan. 2.10 UT by KAIT \citep{2014ATel.5721....1Z}. The coordinates of this SN are R.A. = 22h37m5s.6, Decl. = $+34^\circ$24\arcmin31.9\arcsec (J2000), located at 19\arcsec.7 east and 24\arcsec.2 south of the center of the host galaxy, as shown in Figure \ref{<img>}. Given the Cepheids distance of NGC 7331, $D = 15.1 \pm 0.7$ Mpc \citep{2006ApJS..165..108S}, the distance from SN 2014C to the host center is about 2.2 kpc.

This transient was identified as a young SN Ib \citep{2014CBET.3777....1K} on Jan. 5.54 by Li-Jiang observatory, Yunnan observatories (YNAO) 2.4-m telescope (LJT; \citealp{2015RAA....15..918F}) with YFOSC (Yunnan Faint Object Spectrograph and Camera; \citealp{2019RAA....19..149W}). Cross-correlation with a library of SN spectra using the ``supernova identification" code (SNID, \citealp{2007ApJ...666.1024B}) shows that SN 2014C matches with SN Ib 1999ex at $t \approx -6$ days after brightness maximum (hereafter variable $t$ denotes the time since $V$-band maximum and $\tau$ denotes the time since the explosion). 

We triggered the LJT follow-up observing campaign for this young SN, especially at a daily cadence in the first month after discovery. The monitoring spans from $t \approx -7$ d to $t \approx +340$ d. High-quality template images taken by the LJT+YFOSC at $\sim\,$2500 days after the explosion are applied to the image subtraction for better photometry. Besides, the UV-optical photometry spanning from $t \approx -7$ d to $t \approx +7$ d of the space telescope Swift-UVOT \citep{2004ApJ...611.1005G,2005SSRv..120...95R} is involved in this paper.

\begin{figure*}[ht!]
\includegraphics[width=17cm,angle=0]{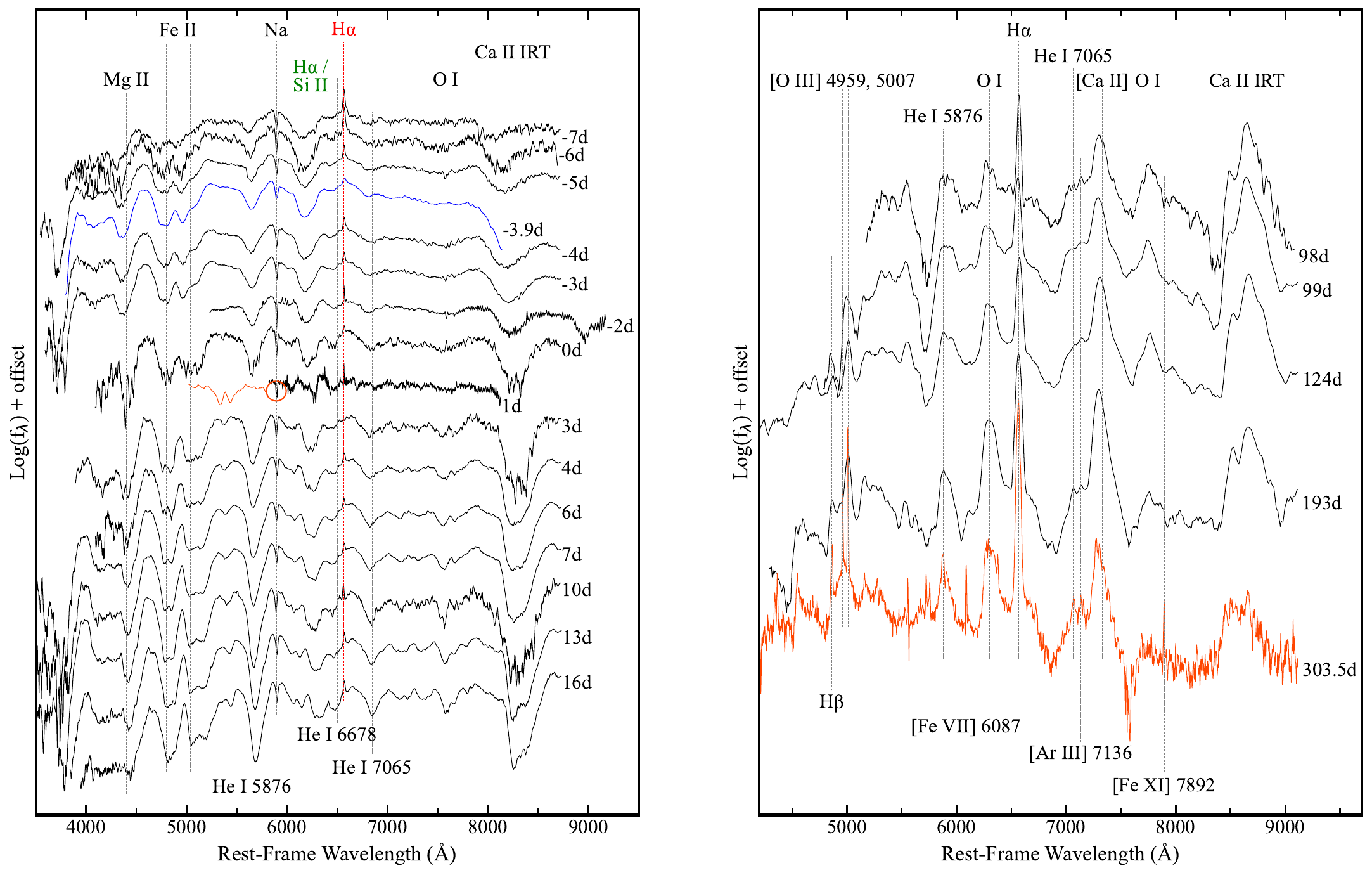}
\centering
\caption{Spectra of SN 2014C obtained by LJT (black), Asiago 1.82 m telescope (blue), and NOT (orange-red). These spectra are shifted vertically for clarity and labeled according to the phase referring to the $V$-band maximum. A bin size = 6 pixels was applied to improve the signal-to-noise ratio of each spectrum. The region around the center wavelength of \NaI, lines at $t \approx 1$ d is zoomed in. The red dotted line marks the H$\alpha$ emission from background contamination.
\label{<14C_Spec>}}
\end{figure*}

\subsection{Photometry}
\label{subsect:photo}
The photometry of SN 2014C is presented in Figure \ref{<14C_LC>}, covering about the first year after the explosion including the photometry of LJT with YFOSC, Asiago 1.82m Copernico telescope with AFOSC (Asiago Faint Object Spectrograph and Camera) and  2.56 m Nordic Optical Telescope (NOT) with ALFOSC (Alhambra Faint Object Spectrograph and Camera). The Johnson-Bessell $UBVRI$-band image of SN 2014C was reduced using the standard procedures of Pyraf \citep{2012ascl.soft07011S}, including corrections for bias, overscan, flat-fielding, and cosmic-ray removal. Twelve local reference stars in the field of this SN are marked in Figure \ref{<img>}. The instrumental magnitudes of these reference stars are converted to the standard $UBVRI$ system in the Vega magnitude by a transformation established by observing a series of Landolt standard stars on some photometric nights. These magnitudes were applied to calibrate the photometry of SN 2014C, as presented in Table \ref{Tab:Pho_Ground2}.

\begin{figure*}[ht!]
\centering
\includegraphics[width=13cm,angle=0]{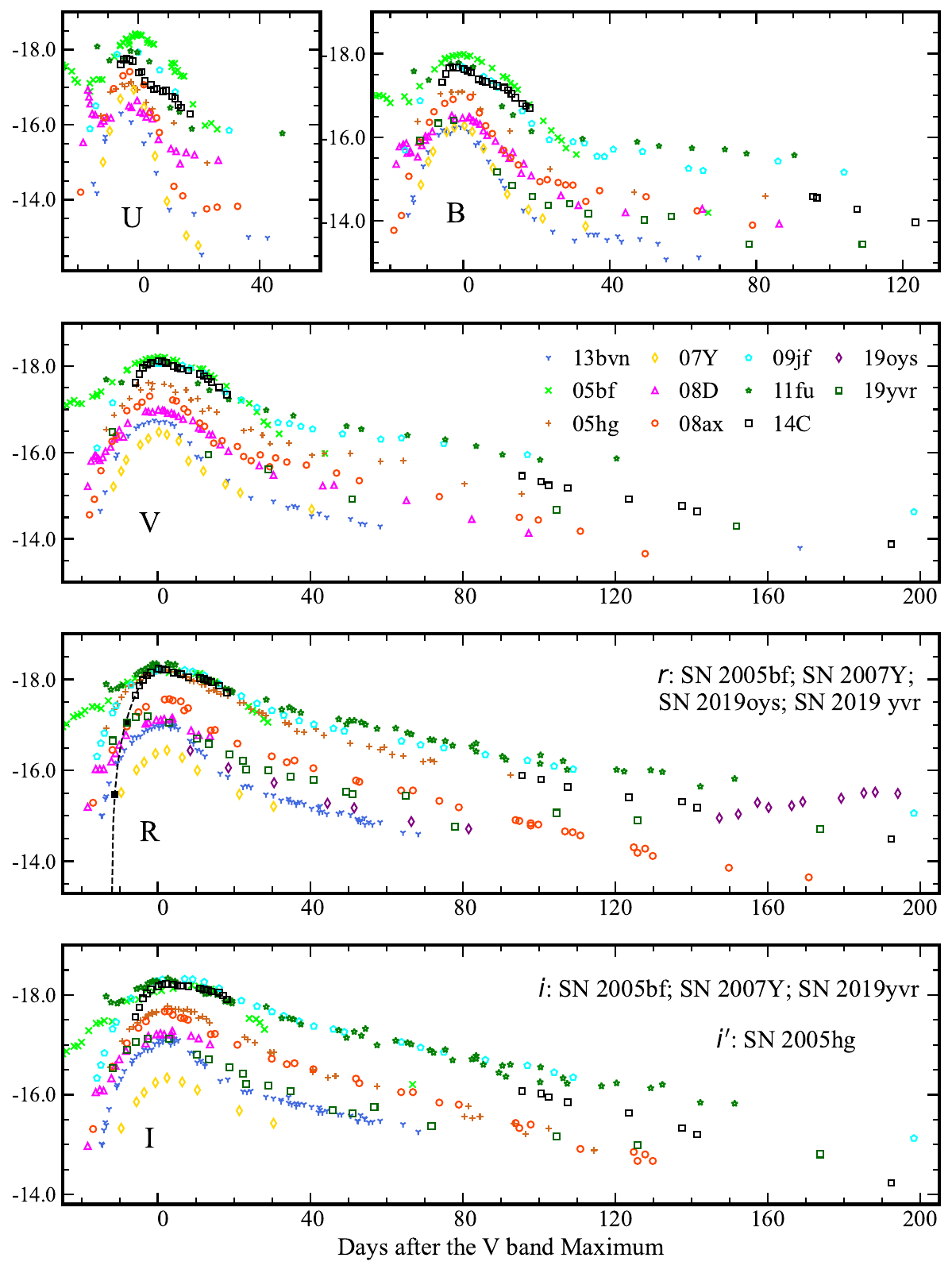}
\caption{$UBVRI$-band absolute light curves of SN 2014C, compared with the light curves of some well-studied SESNe. The two filled squares mark photometry from unfiltered KAIT images obtained on Jan. 5.09 (MJD 56659.10) and from earlier red magnitudes on a pre-discovery image taken on Jan. 2.10 (MJD 56662.09; \citealp{2014ATel.5721....1Z}). A power law fit (dashed line) is used with this photometry (filled squares) and the first three $R$-band measurements. The absolute magnitudes have been temporally shifted to the epoch of maximum brightness in the $V$ band. \label{light_curves}}
\end{figure*}

The photometry of SN 2014C obtained with the $Swift$-UVOT in one UV filter ($uvw1$) and three broadband optical filters ($u$, $b$, and $v$) is presented in Table \ref{Tab:Swiftpho}. The photometry presented here was reduced using the $Swift$ Optical/Ultraviolet Supernova Archive (SOUSA; \citealp{2014Ap&SS.354...89B}) reduction, which includes subtraction of the host galaxy flux using $Swift$-UVOT observations from 2007 May and 2013 April. Table \ref{Tab:Swiftpho} lists the final UVOT UV/optical magnitudes of SN 2014C. The optical magnitudes derived by SOUSA match the $UBV$-band magnitudes obtained by LJT on scales smaller than 0.05 mag. However, the $Swift$-UVOT photometry in \cite{2017ApJ...835..140M} is $\sim$ 0.2 - 0.6 mag brighter than that of SOUSA in the $u-$, $b-$, and $v-$bands, and is $\sim$ 1.0 mag in the $uvw1-$band.  

The discrepancy between the two datasets exhibits an inverse proportionality to both the observation wavelength and the brightness of the SN, which is commonly associated with measurement errors due to host background contamination. The measurements reported in \cite{2017ApJ...835..140M} did not incorporate template subtraction, resulting in an overestimated brightness. Around peak brightness, the difference in the optical bands narrows to approximately 0.2 mag. Therefore, while significant discrepancies exist during the early stages, the two measurements of peak luminosity are generally consistent.

\subsection{Spectroscopy}

Figure \ref{<14C_Spec>} shows spectra of SN 2014C obtained by LJT (+YFOSC), Asiago 1.82 m telescope (+AFOSC; classification spectrum presented in \citealp{2014CBET.3777....1K}) and NOT (+ALFOSC)  covering about 300 days since Jan. 05, 2014. Observational journals for these spectra are listed in Table \ref{Tab:Spec_log}. All of these spectra were calibrated in both wavelength and flux and were corrected for telluric absorption and redshift. The flux of the continuum was then cross-checked with the synthetic photometry computed using the Johnson-Bessell passbands. 

\subsection{Reddening}
\label{subsect:Extin}

The early spectra of SN 2014C exhibit significant absorption of narrow \NaI\ from the host galaxy. For instance, the mid-resolution spectrum at $t \sim +1$ day reveals resolved doublets of sodium, indicating substantial line-of-sight reddening towards SN.

Despite the notable scatter in the data (e.g., \citealp{2011MNRAS.415L..81P, 2013ApJ...779...38P}), the equivalent width (EW) of \NaI\, absorption serves as a tool to roughly estimate the reddening based on empirical correlations. For example, correlations such as $E(B-V) = 0.16\,EW_{\rm Na} - 0.01$ \citep{2003fthp.conf..200T}, and $E(B-V) = 0.25\,EW_{\rm Na}$ \citep{1990A&A...237...79B} can be applied. The measured EW (Na I D) of SN 2014C was 3.19 $\pm\,0.08$ \AA\, corresponding to an average color excess of $E(B-V)_{\rm host}$ = 0.65 $\pm$\,0.20 mag following these relations. This value aligns closely with the estimation provided by \cite{2015ApJ...815..120M}, which yields $E(B-V)_{\rm host}$ = 0.67 mag using the same methodology.

Based on a broader sample of SNe Ib/c, \citet{2011ApJ...741...97D} proposed estimating host-galaxy reddening photometrically. They found that the $V-R$ color of extinction-corrected SNe Ib/c clusters tightly around 0.26 $\pm$ 0.06 mag at $t \approx+10$ days after the $V$-band maximum, and $0.29 \pm 0.08$ mag at $t \approx +10$ days after the $R$-band maximum. This method yields an estimate of $E(B-V)_{\rm total} = 0.67 \pm 0.05$ mag for SN 2014C, assuming an $R_V = 3.1$ Milky Way extinction law for the host galaxy. Considering the Galactic extinction $E(B-V)_{\rm Gal} = 0.08\pm0.01$ \citep{2011ApJ...737..103S}, the host reddening derived from the $V-R$ color is $E(B-V)_{\rm host} = 0.59 \pm 0.05$ mag.

Considering the results of \NaI\,absorption and $V-R$ color, the average value of host galaxy reddening is $E(B-V)_{\rm Host} = 0.62 \pm 0.10$ mag, and the total extinction adopted for subsequent calculations is $E(B-V) = 0.7 \pm 0.1$ mag.

\begin{table}
\centering
\scriptsize
\caption{$UBVRI$-band parameters of SN 2014C}
\begin{tabular}{cccccc}
\hline\hline
Band   &$t_{max}$      & $t_{rise}$  &  $m_{peak}$  & $M_{peak}$ & $\Delta m_{15}$      \\
       &(MJD)          &  (d)        &  (mag)       & (mag)      & (mag)   \\
\hline
$U$	   &56666.48       &   7.57      & 16.54        &  -17.76    &  1.00   \\
$B$    &56668.62       &   9.71      & 16.05        &  -17.70    &  0.62   \\
$V$	   &56669.97       &   11.47     & 14.95        &  -18.11    &  0.48   \\
$R$	   &56671.04       &   12.13     & 14.34        &  -18.22    &  0.33   \\
$I$	   &56672.87       &   13.96     & 13.79        &  -18.21    &  0.26   \\
\hline
\hline
\end{tabular}

\label{Tab:pho_results}
\end{table}

\begin{figure*}[ht!]
\centering
\includegraphics[width=13cm,angle=0]{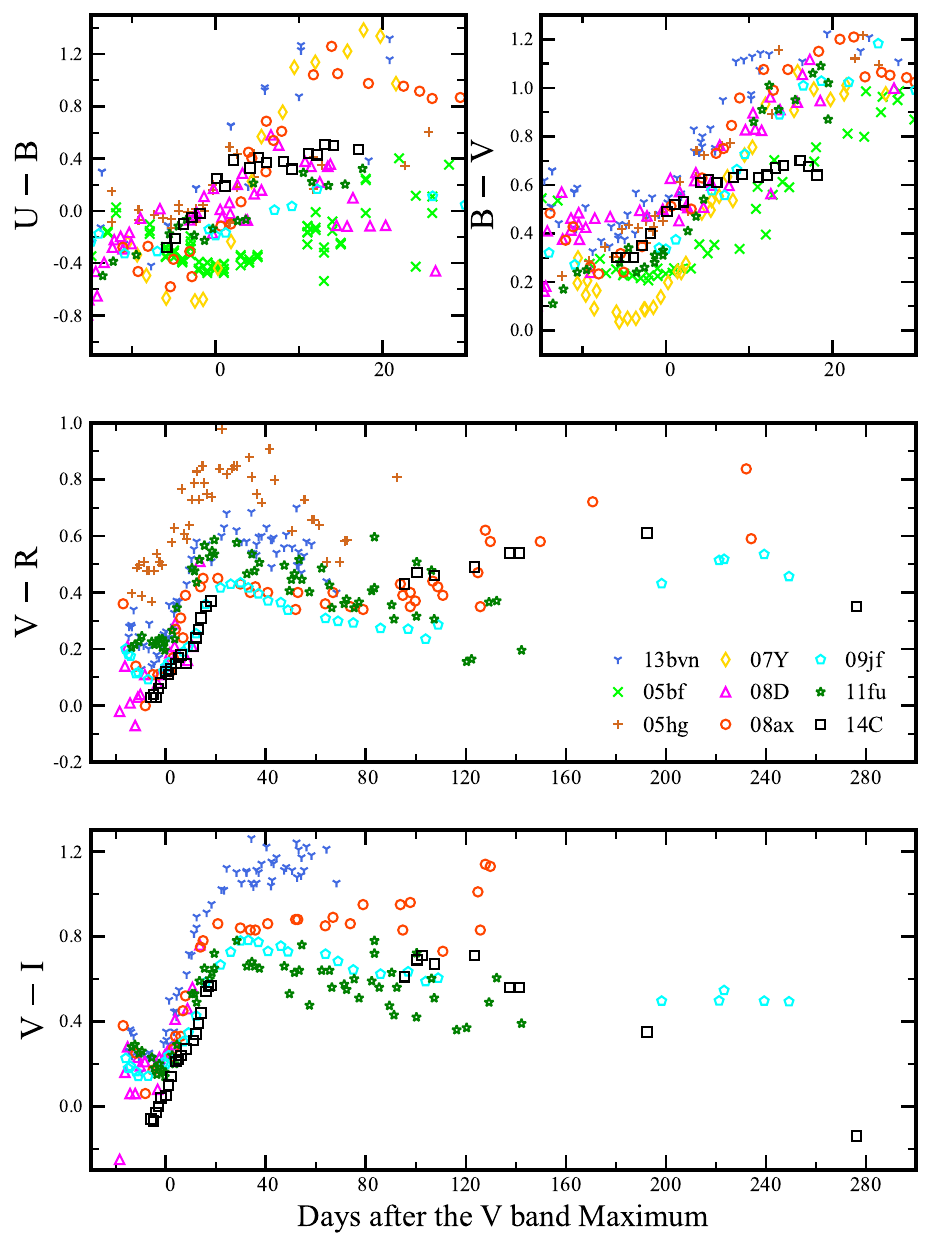}
\caption{Color curves of SN2014C comparing with other SNe Ib/c. All of these samples were corrected with reddening.}
 \label{color_curves}
\end{figure*}

\section{photometric analysis} \label{lc&cc}

\subsection{Light Curves}
Figure \ref{<14C_LC>} shows the photometric evolution of SN 2014C in the first year after the explosion. The light curve parameters of SN 2014C, estimated by low-order polynomial fitting, are presented in Table \ref{Tab:pho_results}. These include the peak time ($t_{\rm max}$), rise time ($t_{\rm rise}$), peak magnitude ($m_{\rm peak}$), absolute magnitude ($M_{\rm peak}$), and post-maximum decline of the light curve in 15 d ($\Delta m_{15}$), are listed in Table \ref{Tab:pho_results}. 

Based on the pre-discovery detection in the clear band from KAIT images and LJT follow-up in the $R$-band, we derived the explosion date of SN 2014C to be MJD = 56658.91 via fireball fitting, as seen in Figure \ref{light_curves}. The fireball model assumes that the photosphere expands uniformly while its surface luminosity remains constant \citep{1982ApJ...253..785A}.
Our estimated explosion date aligns well with the phase via SNID fitting of our identification spectrum. However, the explosion time inferred using the fireball model may have limitations for SNe Ib/c. For instance, some SNe Ib/c, such as SNe 2005bf and 2008D (Figure \ref{light_curves}), exhibit two peaks in their early light curves, potentially attributed to shock breakout and subsequent cooling effects. Due to the absence of sufficiently close and deep non-detection data before the first detection of SN 2014C, its actual explosion time may precede the fireball model estimates. With available data and assumptions about typical supernova evolution, \cite{2017ApJ...835..140M}  estimated MJD = 56656 as the explosion date. \cite{2013ApJ...769...67P} discuss additional uncertainty in estimating the time of explosion from a simple extrapolation of the light curve.

Besides SNe 2005bf and 2008D, Figure \ref{light_curves} also displays the $UBVRI$-band light curves comparisons with other well-sampled SNe Ib/c, including two SN 2014C-like metamorphic SNe 2019oys and 2019yvr, both displayed SN-CSM interaction signatures in their nebular spectra. The $r$- and $i$-band light curves of SNe 2005bf, SN 2007Y and SN 2019yvr, and the $i'$-band light curve of SN 2005hg are plotted in this figure due to the lack of observations in Johnson $RI$-bands. Table \ref{Tab:compared sam} lists the $R/r$-band parameters of these samples. 

SN 2014C reaches its $R$-band peak approximately 12 d after the explosion, which is faster than the rise times of the other SNe Ib/c. To minimize the influence of the shock cooling phase (as observed in SNe 2008D and 2005bf) and focus on the luminosity increase driven by radioactive decay and photospheric expansion, we computed the duration for the luminosity to ascend from half of its peak value to the peak. Even among all the samples listed in Table \ref{Tab:compared sam}, SN 2014C exhibited the swiftest rise, attaining peak luminosity from half its maximum brightness in $t^{R}_{1/2} \approx 7.3$ d, half the typical rise time for SNe Ib.

The rise time of SN depends on factors such as the ejecta mass and explosion energy, with a fast rise potentially indicating a high ratio of explosion energy to ejecta mass (e.g., \citealp{1994Natur.371..227N}). Additionally, the width of the light curve is sensitive to the photon diffusion time relative to the explosion kinetic energy and ejecta mass. A small progenitor radius at the time of explosion can lead to a rapid rise due to a short dynamical time scale. Moreover, a small amount of $^{56}$Ni synthesized in the explosion can also result in a fast rise if the light curve is primarily powered by radioactive decay, as observed in SNe Ia \citep{1982ApJ...253..785A}.

However, SN 2014C exhibits a relatively higher peak luminosity, e.g., $\rm M^V_{\rm max} \approx - 18.1$ mag, compared to the average value of SNe Ib, e.g., $\rm M^V_{\rm max} \approx - 17$ mag \citep{2018A&A...609A.136T}. As depicted in Figure \ref{light_curves}, among the comparison samples, only the peculiar Ib/c SN 2005bf shows a brighter peak than SN 2014C in the $U$ and $B$ bands, while they are similar in the $VRI$ bands. Additionally, SN 2011fu and SN 2009jf display similar peak brightness to SN 2014C, but they exhibit a brighter nebular phase than SN 2014C.

Among these SNe, the decline of SN 2005bf is faster due to the progenitor losing almost all of its envelope before exploding, with no observed interaction between the ejecta and the preceding slow stellar wind. The slow decline rate of the light curve in SN 2014C, SN 2009jf, and SN 2011fu suggests some form of interaction or additional energy sources. Nevertheless, SN 2014C declines more slowly than the comparison samples. A plateau-like structure can be observed in the $BVRI$ band from 10 to 20 days after the peak brightness. Suppose this plateau is caused by the kinetic energy of the ejecta being transferred due to SN-CSM interaction. In that case, a hydrogen envelope shell might exist at a distance from SN of $2 \times 10^{15}$ cm to $3 \times 10^{15}$ cm, assuming an expanding velocity of 10,000 km/s. 

Between $100 < t < 200$ d, the decline rate of SN 2014C is faster than expectations from radioactive decay of $^{56}$Co, suggesting the presence of $\gamma$-ray leakage during this period. Conversely, at $t \gtrsim 200$ d, the light curve decline rate decelerates relative to the radioactive decay rate, implying additional energy sources beyond the decay of $^{56}$Co. As shown in Figure \ref{light_curves}, despite significant SN-CSM interaction signals observed in its nebular spectra, the luminosity of SN 2014C at $t < 200$ d  did not decline significantly more slowly than that of other SNe Ib relative to their peak brightness. During a similar timeframe, SNe 2019oys and 2019yvr exhibited a more gradual decline in their light curves, with SN 2019oys even experiencing rebrightening at $t > 140$ d. The diversity among these metamorphic SNe suggests variations in the efficiency of converting shock kinetic energy into radiation through SN-CSM interaction.

\begin{figure*}
\centering
\includegraphics[width=15cm,angle=0]{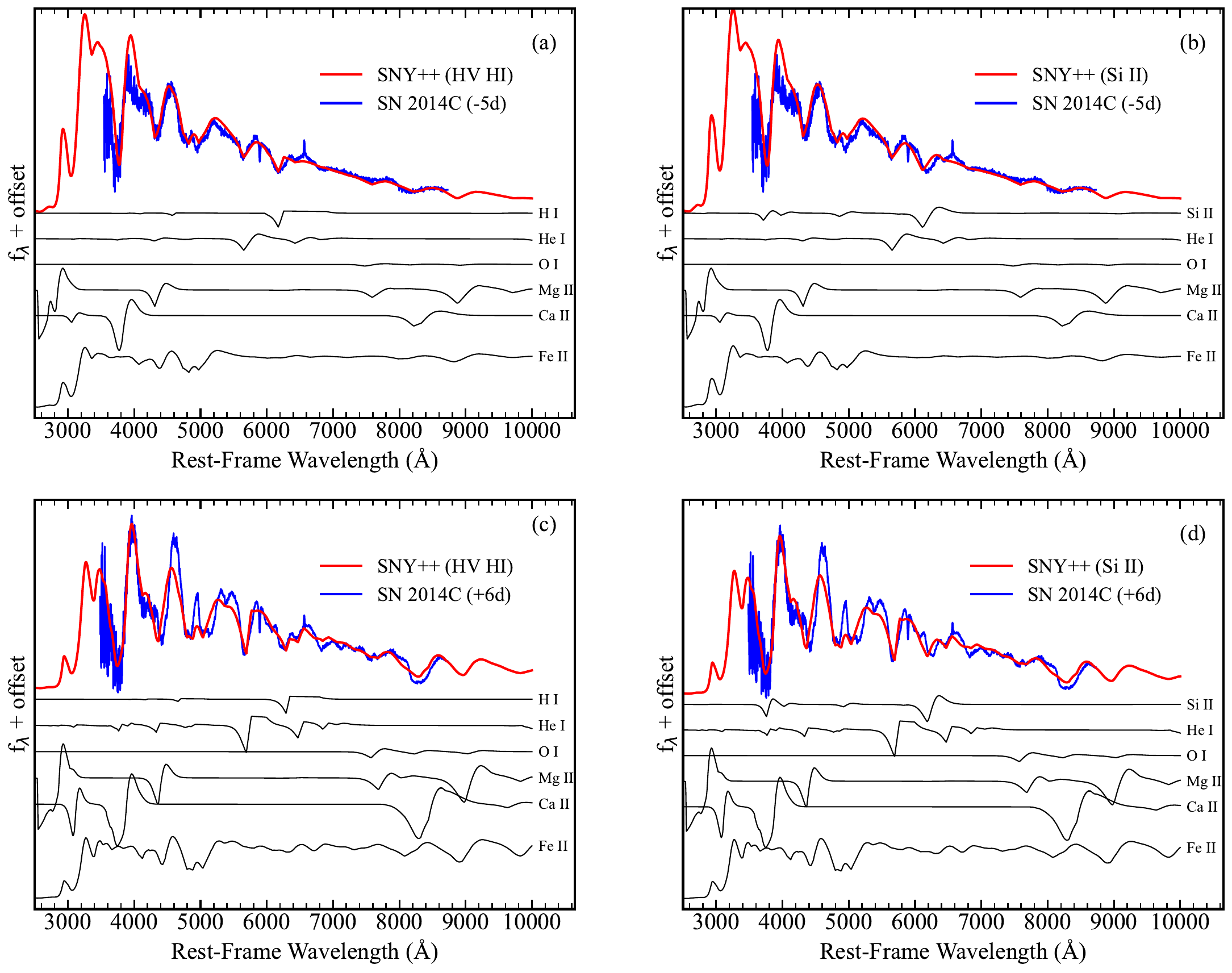}
\caption{Spectra of SN 2014C at -5 d and +6 d reproduced by SYN++, including the element of H (panel a and c) or Si (panel b and d), respectively.}
\label{syn}
\end{figure*}

\subsection{Color Curves}

Figure \ref{color_curves} illustrates the color curves of SN 2014C alongside some comparison samples. All color curves have been corrected for reddening using the $E(B-V)$ values listed in Table \ref{Tab:compared sam}, assuming an extinction law with $R_V = 3.1$. In this figure, SN 2014C exhibits a relatively red color in $U-B$ and $B-V$, while displaying a blue color in $V-R$ and $V-I$, suggesting differences in energy distribution and temperature evolution.

In the early phase, contrary to the `U-turn' profile observed in most comparison samples, all color curves of SN 2014C show a monotonically decreasing trend from blue to red, indicating a decrease in temperature for this SN at $t \lesssim 5$ days. Between $t \sim 5 - 20$ days, the $U-B$ and $B-V$ color curves of SN 2014C stabilize, while its $V-R$ and $V-I$ colors appear bluer compared to the comparison sample. This observation may be linked to the plateau-like structure observed in the light curves during this phase. Furthermore, at later times ($t \gtrsim 120$ days), the nebular $V-I$ color of SN 2014C becomes even bluer compared to its earlier state. This change can be attributed to an increased flux of H$\alpha$ emissions resulting from interactions between SN and CSM, as evident from nebular spectra.

\section{Spectra Analysis}
\label{spectra}

\subsection{First month}
In the left panel of Figure \ref{<14C_Spec>}, densely sampled spectra reveal clear blueshifted \HeI\, $\lambda\lambda$ 5876, 7065 lines in the first month after the explosion, classifying SN 2014C as an SN Ib \citep{2008Sci...321.1185M, 2008Natur.453..469S}. Besides, these spectra are dominated by the bellowing features: (1) absorption between 4400 and 4600 \AA\,attributed to \MgII; (2) two absorption lines at $\sim$ 5000 \AA\,caused by \FeII; (3) a gradually strengthening \OI\,$\lambda$\,7774 absorption; (4) the blended \CaII\,$\lambda\lambda$ 8498, 8542, 8662 triplets at the red-end.

An absorption feature around 6200 \AA\,may arise from H$\alpha$ or \SiII\,$\lambda$ 6355. 
\cite{2015ApJ...815..120M} proposed that this feature stems from high-velocity H$\alpha$ and successfully reproduced the spectrum at $t \approx -4$ d within the elementary supernova spectrum synthesis code (SYN++; \citealp{2011PASP..123..237T}). Nonetheless, it is crucial to acknowledge that their analysis was limited to a single early spectrum, precluding further inferences based on the evolution of this absorption feature. However, as illustrated in Figure \ref{syn}, both high-velocity H$\alpha$ and the photospheric \SiII\,$\lambda$ 6355 can fit the pre-maximum spectra within SYN++. The origin of this absorption is better constrained by examining multiple early spectra and tracing its evolutionary path.

\begin{figure}
\centering
\includegraphics[width=8cm,angle=0]{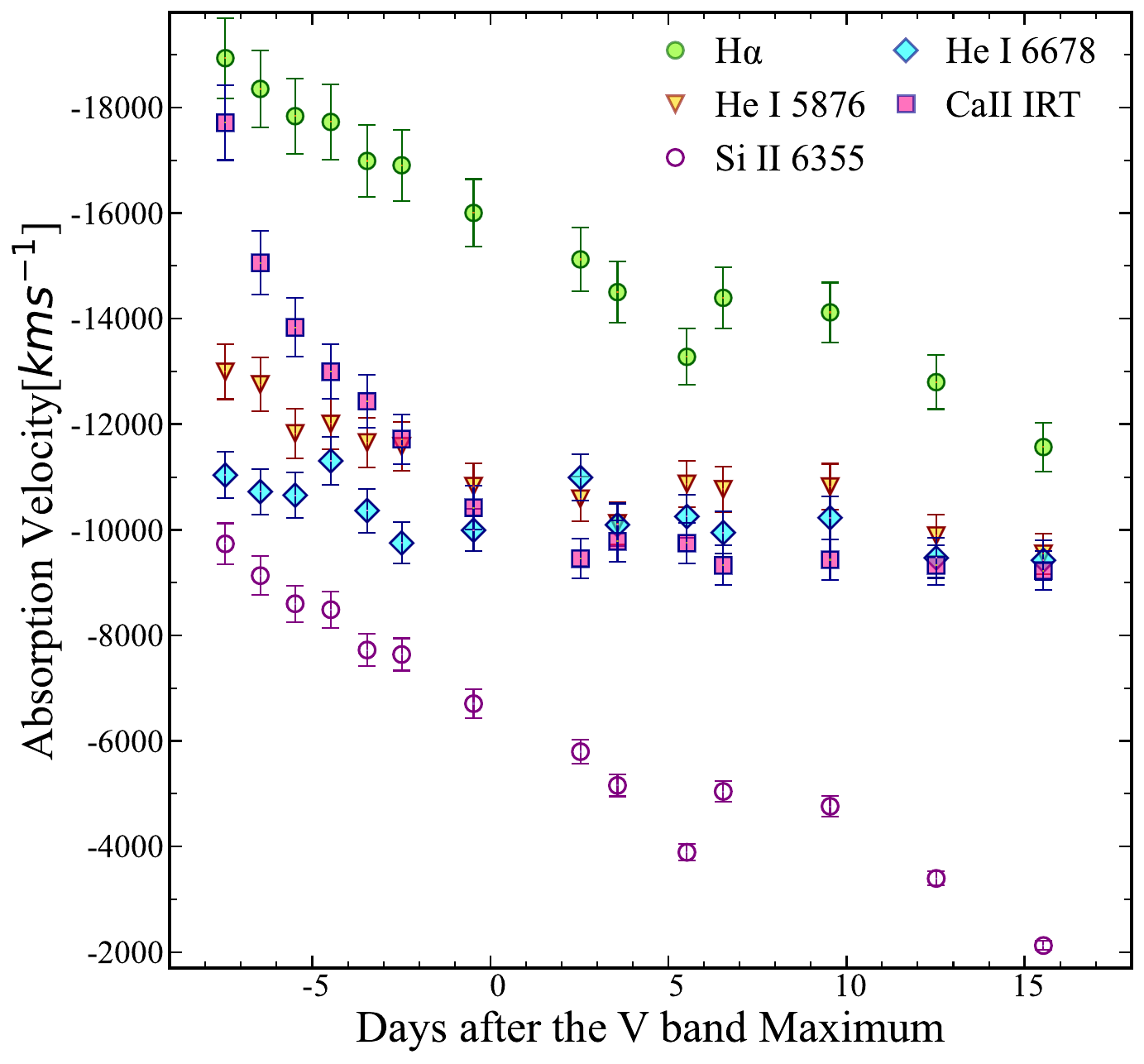}
\caption{Evolution of line velocities for H$\alpha$, \SiII\,$\lambda$ 6355, \HeI\,$\lambda$5876, \HeI\,$\lambda$6678, and \CaII\,IR triplet of SN 2014C, derived from the absorption minima of each spectral line. Note: velocities of H$\alpha$ and \SiII\,$\lambda$ 6355 are derived from the same absorption line around 6200\AA.}
\label{vels}
\end{figure}

\begin{figure}
\centering
\includegraphics[width=8cm,angle=0]{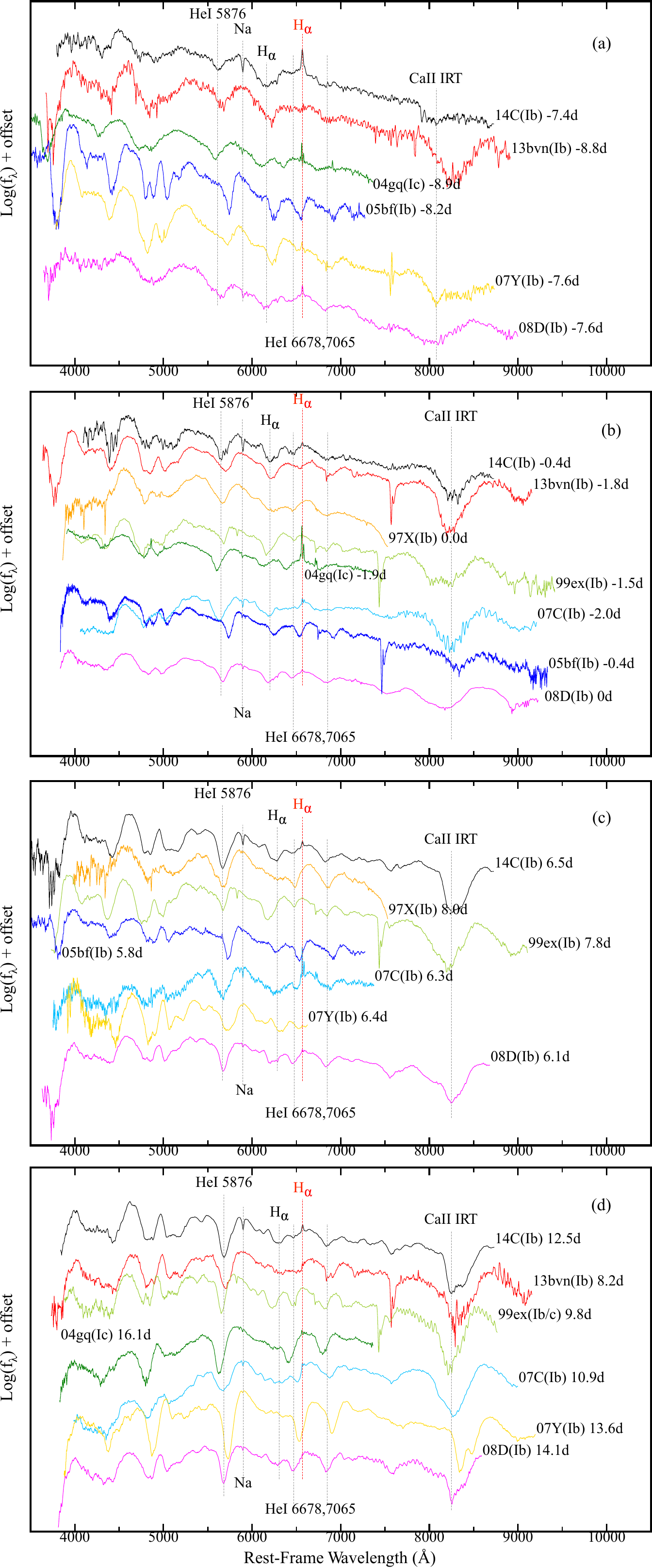}
\caption{Comparison of spectra of SN 2014C with similar epoch spectra of other well-studied SNe. Labels in parentheses to the right of each spectrum indicate days from the $V$ band maximum time. References for each spectrum are listed in Table \ref{Tab: spec}. These spectra have been corrected for redshift and reddening.}
\label{comp}
\end{figure}

\begin{figure}
\centering
\includegraphics[width=8cm,angle=0]{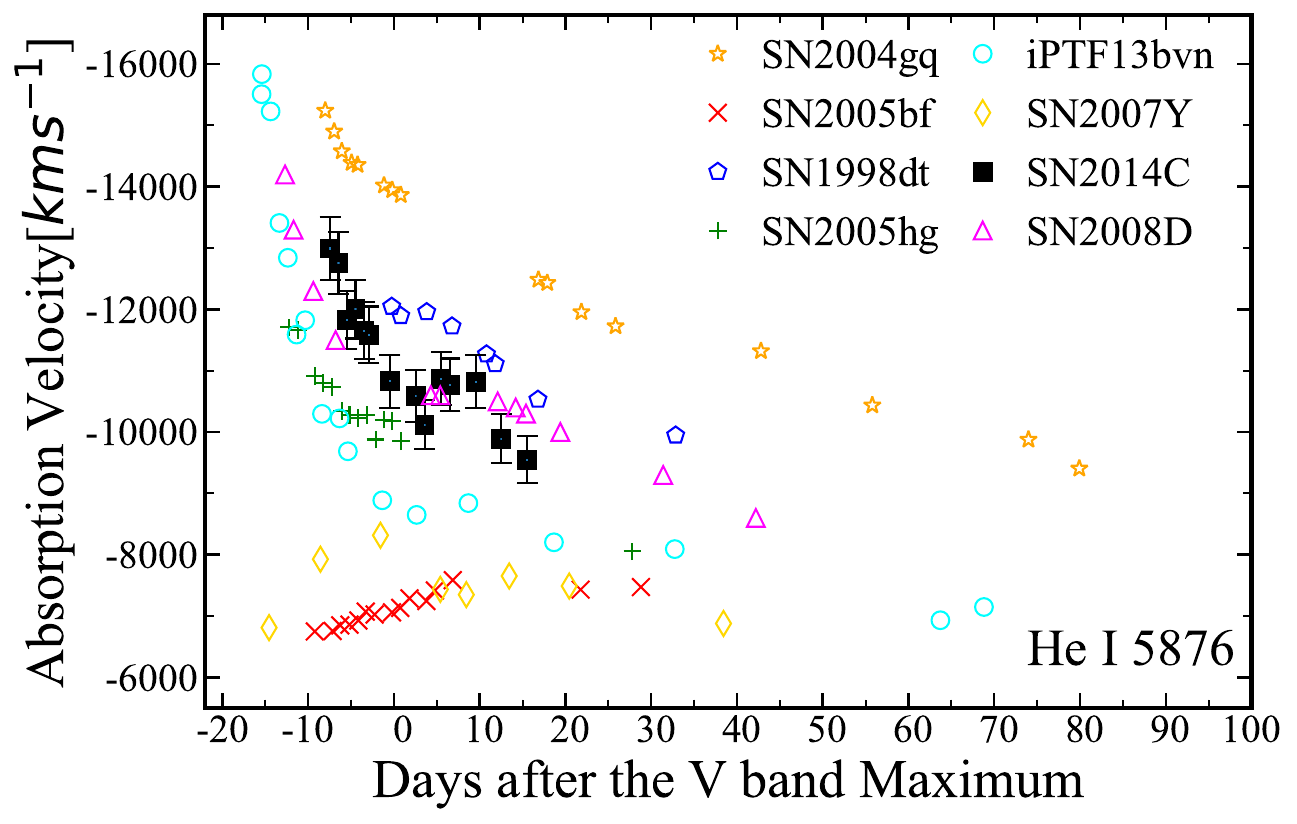}
\caption{Velocity derived from the absorption minimum of \HeI\,$\lambda$ 5876\AA, including SN 1998dt \citep{2001AJ....121.1648M}; SN 2005bf \citep{2005ApJ...633L..97T}; SN 2004gq, SN 2005hg and SN 2008D (Fig. 13 of \citealp{2009ApJ...702..226M}); iPTF13bvn \citep{2014A&A...565A.114F}, SN 2007Y \citep{2009ApJ...696..713S}, and SN 2014C (this work).}
\label{vel_5876}
\end{figure}

Figure \ref{vels} illustrates the velocity evolution of this absorption component under the assumptions of H$\alpha$ or \SiII\,$\lambda$ 6355, comparing it to the velocities of \HeI\,and \CaII. These velocities decrease over time due to the decreasing material velocity as the SN ejecta undergoes homologous expansion. If the component around 6200\AA\, is \SiII\,$\lambda$ 6355, its velocity in the first two days is slightly lower than that of \HeI\,$\lambda$ 6678. The variation of optical depth can account for this minor velocity difference. However, as time progresses, the velocity of this component decreases rapidly, clearly distinguishing it from other photospheric components. At $t \sim 16$ d, assuming it is \SiII$\lambda$ 6355, its velocity is only $\sim$20\% of that of He and Ca, making this absorption unlikely to be photospheric Si. This rapid decline in velocity aligns with the characteristics of high-velocity features consistent with the high-velocity H$\alpha$ assumption. As shown in panels c and d of Figure \ref{syn}, the  assumption of H$\alpha$ provides a better fit for the spectrum at $t\approx +6$ d compared to \SiII\, $\lambda$ 6355. Therefore, we confirm that the absorption near 6200\AA\,is due to high-velocity H$\alpha$ rather than \SiII $\lambda$ 6355 based on velocity evolution.

The higher velocity of H$\alpha$ absorption compared to \HeI\,and \CaII\ suggests it originates in the hydrogen at the outer layer of the ejecta. At $t \sim 16$ d, the velocity of H$\alpha$ becomes comparable to that of other photospheric components, indicating an extensive distribution of hydrogen that spans both the outer and inner layers of the ejecta.

Through SNID fitting, we identified a sample group with early spectra similar to SN 2014C, as presented in Figure \ref{comp}. These samples are classified as SNe Ib or Ib/c in the literature. Generally, the spectra of SN 2014C are more similar to SN 2008D and iPTF 13bvn than other comparisons. They exhibit similarities in the profiles and evolutionary trends of \HeI\, line, and high-velocity H$\alpha$. Their main differences are reflected in the velocity and width of the \CaII\,IRT. SN 2014C exhibits a faster and narrower \CaII\,IRT velocity at $t\approx -7$ d, but around two weeks after maximum light, its \CaII\,IRT becomes stronger instead. Those reflect the similarities and differences in the ejecta structures of these SNe.

Notably, early high-velocity H$\alpha$ components similar to SN 2014C are prevalent in the early stages of these SNe Ib/c. However, the high-velocity H$\alpha$ absorption feature of SN 2014C persists for a relatively long duration, with significant absorption still evident at $t \sim 16$ d. In contrast, the corresponding absorption components in iPTF 13bvn, SN 2004gq, SN 2007C, and SN 2007Y diminish around ten days after maximum light. The slower evolution of H$\alpha$ in SN 2014C compared to these companions suggests a denser stripped H-rich envelope in SN 2014C relative to the other samples.

Figure \ref{vel_5876} illustrates the velocities of \HeI\,$\lambda$ 5876 \AA\, the strongest optical \HeI\, line in SN 2014C, and comparison sample. \HeI\, lines result from nonthermal excitation and ionization of gamma-ray photons produced during the radioactive decay of $^{56}$Ni and $^{56}$Co \citep{1987ApJ...317..355H, 1991ApJ...383..308L, 1998MNRAS.295..428M}, revealing variations in the ejecta of SNe Ib, which exhibit a wide velocity range. The velocities of SN 2014C closely resemble those of SN 2008D, following a similar evolutionary trend to iPTF13bvn, suggesting comparable ionization states and optical depths of the ejecta during the photospheric phase.

\begin{figure}
\centering
\includegraphics[width=8.5cm,angle=0]{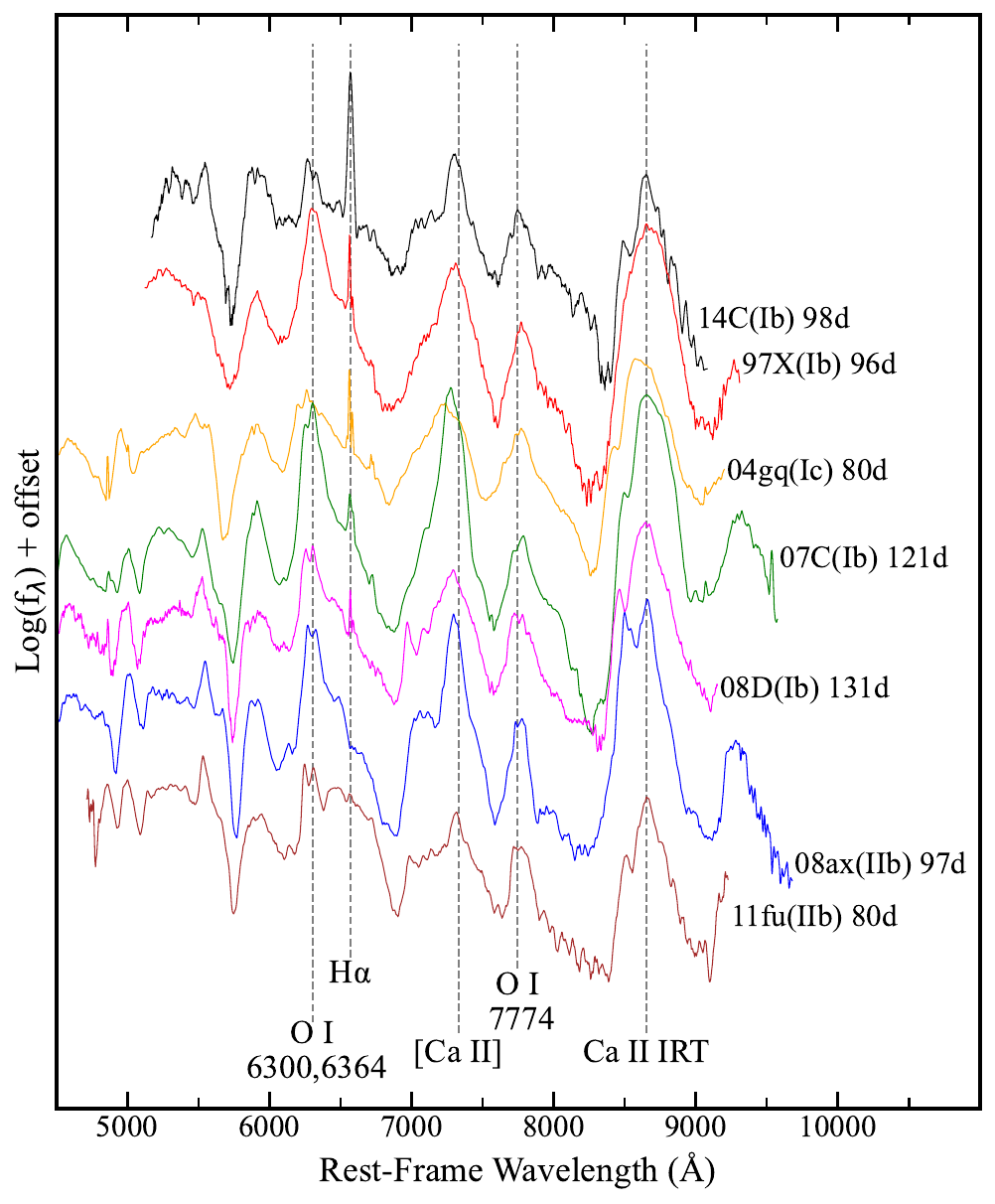}
\caption{Comparison of nebular phase spectra of SN 2014C with other similar samples or well-studied SNe including SN 1997X \citep{2009MNRAS.397..677T}, SN 2004gq \citep{2014AJ....147...99M}, SN 2007C \citep{2009MNRAS.397..677T}, SN 2008D \citep{2009ApJ...702..226M}, SN 2008ax \citep{2011MNRAS.413.2140T}, SN 2011fu \citep{2015MNRAS.454...95M}. Labels in parentheses to the right of each spectrum indicate days from the $V$ band maximum time. These spectra have been corrected for redshift and reddening.}
\label{neb_spec}
\end{figure}

\begin{figure}
\centering
\includegraphics[width=6.5cm,angle=0]{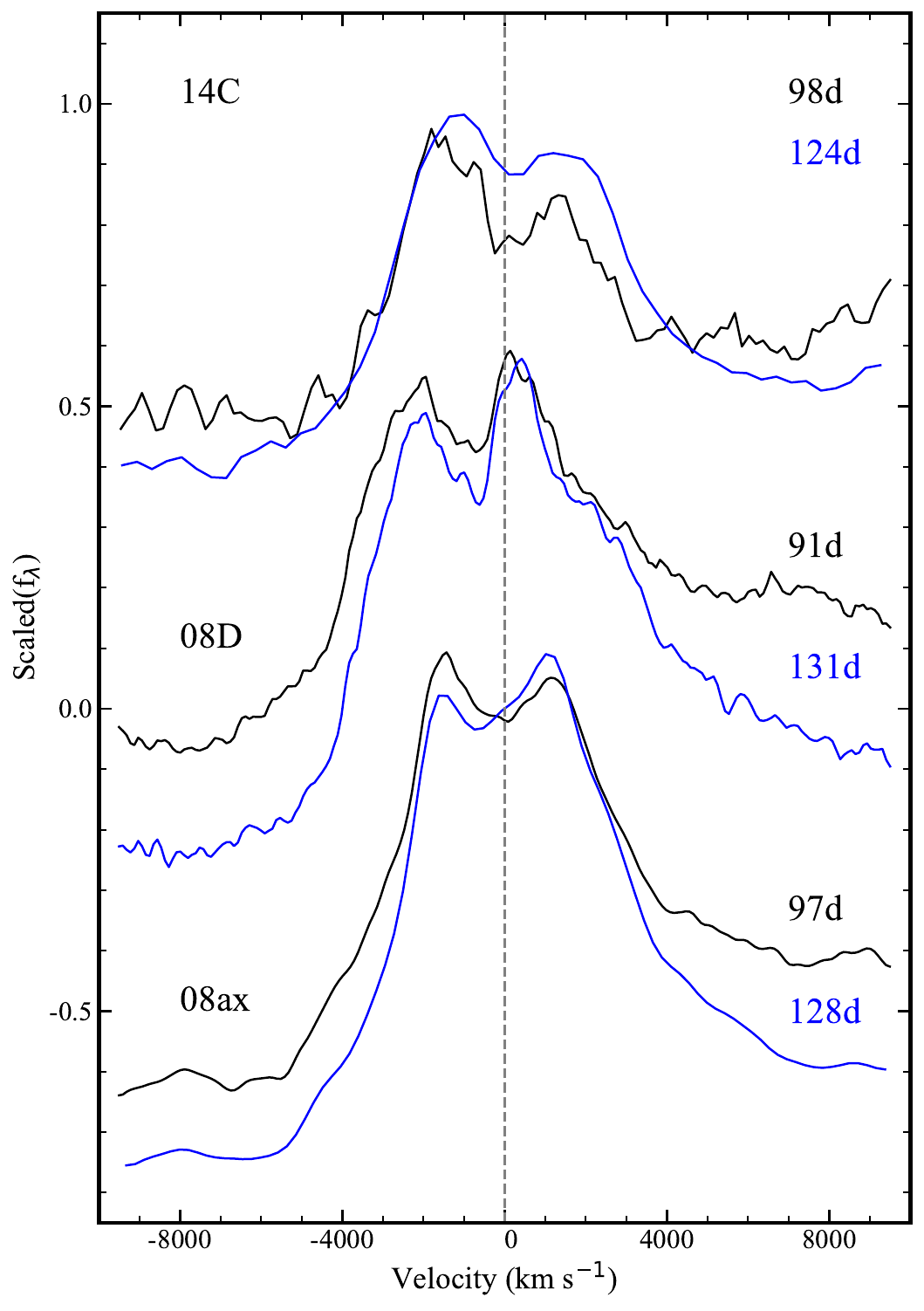}
\caption{The emission lines of [\OI] 6300, 6364 double peak of SNe 2014C, 2008D \citep{2009ApJ...702..226M}, 2008ax \citep{2011ApJ...739...41C, 2011MNRAS.413.2140T} in velocity space. The zero point is 6300\AA\,and is indicated by the gray dashed line. These spectra were corrected for the redshift. }
\label{Oxygen_compare}
\end{figure}

\subsection{Nebular Phase}

SNe typically become fully transparent around 100-200 days after maximum light, though some may take up to a year to complete this transition \citep{2004ApJ...614..858M}. During this period, the ejecta becomes optically thin, offering a clearer view of the explosion's core, a stage known as the nebular phase. In this phase, emission lines overlay the residual photospheric spectrum.

The right panel of Figure \ref{<14C_Spec>} displays five nebular spectra of SN 2014C taken from approximately $ t \approx 98$ d to $t\approx$ 303 d. These spectra are dominated by strong forbidden emission lines of intermediate-mass elements (e.g., [\OI] $\lambda$$\lambda$\,6300,6364 and [\CaII] $\lambda$$\lambda$\,7292,7324). Unlike the narrow H$\alpha$ observed in the early phase, stronger H$\alpha$ emission with moderate width (e.g., $\sim$ 1000 \kms) in the nebular spectra indicates significant SN-CSM interaction. This is observed approximately 18 days earlier than reported by \cite{2015ApJ...815..120M}. Given a velocity of $\sim$ 10000 \kms\, the inner radius of the CSM is estimated to be $\sim 10^{16}$ cm. The presence of strong hydrogen emission lines at this epoch suggests that the primary coolant might be the hydrogen of the stripped envelope \citep{1987ApJ...322L..15F, 1986ApJ...308..685U}.

\begin{figure*}
\centering
\includegraphics[width=15cm,angle=0]{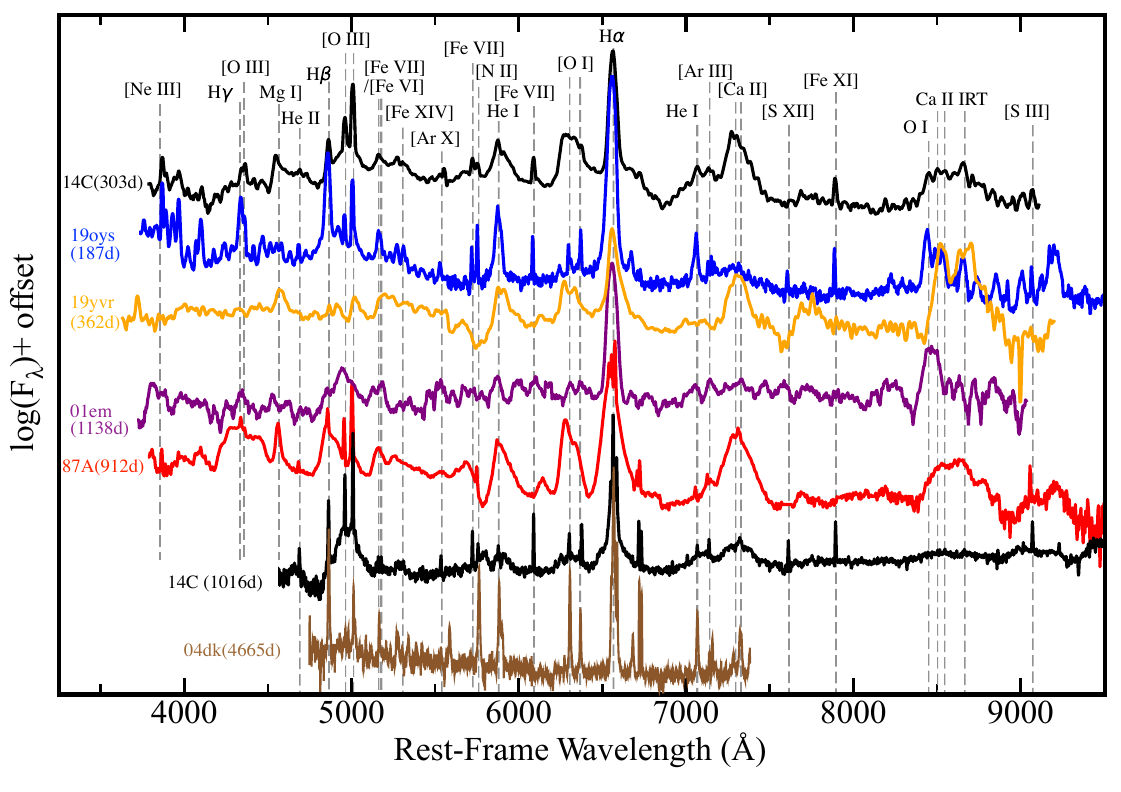}
\caption{Spectra of SN 2014C at $t\sim$ 303 d (this work) and $t \sim 1016$ d \citep{2019MNRAS.482.1545S} comparing to that of SN 2014C-like SNe 2019oys \citep{2020A&A...643A..79S}, 2019yvr  \citep{2024MNRAS.529L..33F},  2001em \citep{2019MNRAS.482.1545S}, 2004dk  \citep{2018MNRAS.478.5050M} and SN II 1987A (from the Padova-Asiago database). Labels in parentheses to the left of each spectrum indicate
days from the peak brightness. These spectra have been corrected for redshift and reddening. Various bin sizes were used for the different spectra, depending on the original signal-to-noise ratio. }
\label{nebu87A}
\end{figure*}

Figure \ref{neb_spec} presents the nebular spectrum of SN 2014C at $t\approx 98$ d alongside spectra from some typical SNe Ib/c. The H$\alpha$\,emission line observed in the SNe 1997X, 2004gq, 2007C, and 2008D spectra is mainly due to the background radiation. Notably, its intensity and width are significantly weaker compared to SN 2014C. In addition to its more pronounced H$\alpha$\,emission line, SN 2014C shares similar nebular traits with its counterparts, suggesting a similar core formed in these explosions. The observed SNe Ib characteristics suggest that the CSM surrounding SN 2014C was either not highly opaque or featured a ring-like structure that allowed light to escape from the inner part of SN.

At $t\approx 98$d, SN 2014C shows double-peaked oxygen emission lines consistent with the comparisons except for SN 1997X. This is due to the intrinsic doublet of  [\OI] $\lambda$$\lambda$\,6300,6364. We do not observe the double-peaked structure in [\OI] $\lambda$\,7774 of SN 2014C, seen in SN 2007C and SN 2008D. This difference in oxygen profile relates to the structure of the ejecta. For example, the double-peaked [\OI] $\lambda$\,7774 might suggest that the oxygen-rich ejecta are arranged within a ring- or torus-like structure viewed along the equatorial plane \citep{2005Sci...308.1284M, 2009MNRAS.397..677T, 2008Sci...319.1220M}.

Figure \ref{Oxygen_compare} highlights the [\OI] $\lambda$$\lambda$6300,6364 emission lines for SNe 2014C, 2008D, and 2008ax, showcasing differences in their intensities and velocities.
Although the intensities of the double peaks in SNe 2014C and 2008ax are inconsistent, the positions of their respective double peaks relative to 6300 \AA\,are essentially the same. This structure is classified as symmetric peaks in the study of [\OI] line profiles by \cite{2010ApJ...709.1343M}. They suggested that the symmetric structure of the [\OI] lines originates from the doublet produced by a single emitting source located at the front of SN, moving toward the observer.
Conversely, the double peaks of SN 2008D are asymmetrically distributed around 6300 \AA\, with the overall profile exhibiting a blueshift and the velocity of the [\OI] $\lambda$\,6364 line nearing zero, indicating a difference in viewing angle or movement of emitting source compared to SN 2014C.

The exhibition of intermediate H$\alpha$ emission lines at $t \gtrsim 100$ d leads some to classify SN 2014C as SN IIn. However, \cite{2015ApJ...815..120M} found that the spectra of SN 2014C at $t > 300$ d did not fully align with the nebular spectra of SNe IIn. For instance, the nebular spectra of SN 2014C have no broad emission component beneath the intermediate H$\alpha$ emission line due to H-rich ejecta, suggesting that its ejecta were H-poor at this time.

Figure \ref{nebu87A} compares the spectra of SN 2014C at $t\approx 303$ and 1016 d with four metamorphic SNe from Ib to IIn, including SNe 2001em \citep{2019MNRAS.482.1545S}, 2004dk \citep{2018MNRAS.478.5050M}, 2019oys \citep{2020A&A...643A..79S}, and 2019yvr \citep{2024MNRAS.529L..33F}, captured during the emergence of intermediate H$\alpha$ emission from SN-CSM interaction. Highly ionized narrow emission lines of Fe, Ar, S, and N seem to be common in these metamorphic SNe. Due to the limitations of spectral resolution, the widths of these narrow lines are often comparable to the instrumental broadening, suggesting their intrinsic widths do not exceed 100 \kms, which is consistent with stellar wind velocities. This implies that these narrow emission lines may originate from the unshocked CSM \citep{2006ApJ...641.1051C}.

The narrow lines of H$\beta$ and [\OIII] $\lambda\lambda$\,4959,5007 can easily be mistaken for radiation from the galactic background. However, in the case of SN 2014C, these lines are superimposed on a broader profile. This resembles the scenario where narrow emission lines generated by photoionization in SN-CSM interactions are superimposed on broad emission lines produced by electron scattering \citep{2017A&A...603A..51D}. During seven-year spectral monitoring, \cite{2022ApJ...930...57T} observed the evolution of the intensity of these two [\OIII] lines. Therefore, these narrow lines are likely to originate from interactions rather than the galactic background.

At $t \approx 303 $, the notable Ib components in SN 2014C, e.g., emissions of [\OI], [\CaII] and \CaII\,IRT at the width of $\sim$ 3000 \kms, indicates that the energy from the SN still contributed to the observed luminosity, albeit this contribution was progressively diminishing. Subsequently, at $t \gtrsim 330$ d, the typical Ib emission lines gradually weakened in SN 2014C only weak [\CaII] and [\OI] emission lines were observable, with most other spectral features being produced by SN-CSM interactions \citep{2018MNRAS.478.5050M}. At $t > 1200$ d, the Ib components were almost undetectable in the spectrum, primarily due to the extremely low luminosity of SN itself \citep{2022ApJ...930...57T}.

Like SN 2014C, the spectrum of SN 2019oys at $t\approx 187$ d exhibits narrow emission lines from highly ionized elements such as Fe, Ne, and Ar. However, the [\CaII] emission lines in SN 2019oys are significantly fainter than those in SN 2014C at $t \approx$ 303 d, and the \CaII\,IRT lines of the former are relatively weak. Additionally, the [\OI] $\lambda\lambda$ 6300, 6364 emissions in SN 2019oys are as narrow as seen in the spectrum of SN 2014C at $t\approx 1016$ d, lacking the broadening typically seen in SNe Ib at about 200 d after the explosion, suggesting that its oxygen emission originates primarily from CSM rather than SN itself. The broad emission lines of O and Ca disappeared or became very weak in the spectrum of SN 2019oys, indicating that the outer CSM heavily obscured the inner region of this SN. Thus, the CSM of SN 2109oys has a significantly different optical depth or structure from that of SN 2014C.

\begin{figure*}
\centering
\includegraphics[width=0.8\linewidth,angle=0]{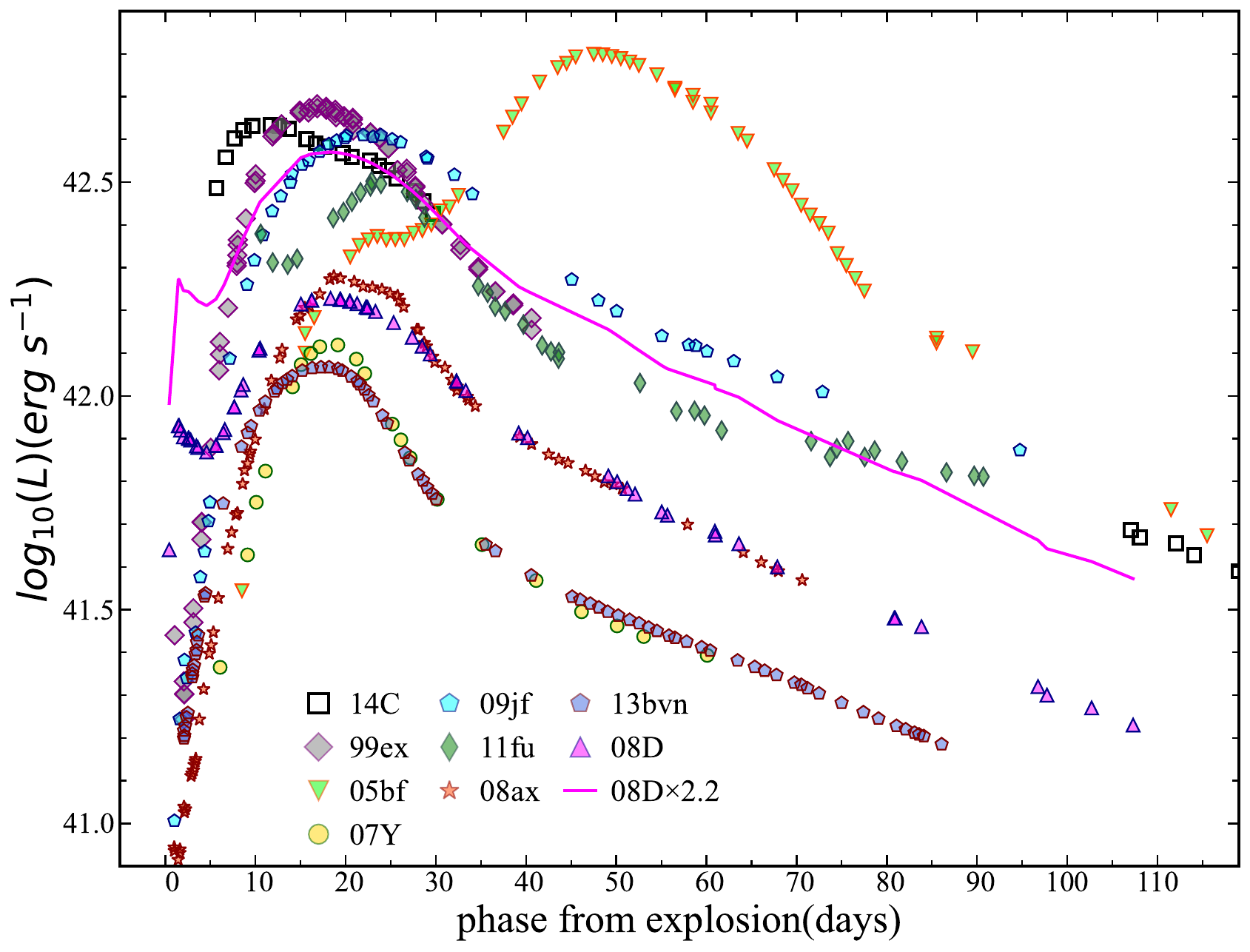}
\caption{The bolometric light curve of SN 2014C compared with some well-studied SESNe, including: SNe IIb SN 2008ax \citep{2008MNRAS.389..955P}, SN 2011fu \citep{2013MNRAS.431..308K}; SN IIb/Ib SN 2007Y \citep{2009ApJ...696..713S}; SNe Ib SN 1999ex \citep{2002AJ....124.2100S}, SN 2008D \citep{2008Sci...321.1185M,2009ApJ...692L..84M}, SN 2009jf \citep{2011MNRAS.416.3138V}, iPTF13bvn \citep{2014A&A...565A.114F} and SNe Ib/c SN 2005bf \citep{2006ApJ...641.1039F}. The light curve of SN 2008D scaled by a factor of 2.2 is plotted in solid-line. }
 \label{bolos}
\end{figure*}

SN 2019yvr exhibits broad lines of Ca, O, and H akin to SN 2014C, yet it lacks the narrow emission lines typical of highly ionized species, such as Fe and Ne. This suggests that, at this stage, the SN core of SN 2019yvr is comparable to that of SN 2014C, but its outer CSM differs.

The first interaction spectrum of SN 2001em, dominated by intermediate H$\alpha$ emission, was observed around 1000 days post-explosion. Due to the low signal-to-noise ratio, no narrow emission lines originating from highly ionized species of Fe, Ar, and O, were discernible in this spectrum. Therefore, it is unclear whether its CSM shares a similar composition and ionization state with SNe 2014C and 2019oys. The relatively weak broad Ca and O emissions imply that the contribution of SN was weaker than the SN-CSM interaction, as observed in SN 2014C at the same phase.

Over 4000 days after the explosion, the first spectrum capturing the interaction between SN 2004dk and its CSM was obtained. By this time, the brightness of SN had diminished, allowing the spectrum to be dominated by intermediate H$\alpha$ emission and narrow ionized lines from elements like He, N, Ar, and Fe. Similar to SNe 2014C and 2019oys, SN 2004dk showed [\FeVII] $\lambda$5158, but did not exhibit the [\FeVII] $\lambda\lambda$\,5720,6087, suggesting similar yet distinct CSM characteristics among these SNe.

Through comparison of Figure \ref{nebu87A}, we found that the spectrum of SN 2014C at $t \approx 303$ d is more similar to that of the type II SN 1987A at $t \approx 912$ d. However, the broad H$\alpha$ and H$\beta$ emissions in the spectrum of SN 1987A, comparable to those of [\OI] and [\CaII], more likely originate from SN itself rather than from SN-CSM interaction. Furthermore, SN 1987A exhibits narrow ionized emission lines such as [\NeIII], [\OIII], and \HeII\,but lacks highly ionized Fe lines, highlighting significant differences in CSM composition compared to SN 2014C. SN 1987A originated from the explosion of a blue supergiant \citep{1989ARA&A..27..629A}, a H-rich star that differs significantly from the typical progenitor of SNe Ib/c. Therefore, the differences in the narrow components of its spectrum compared to SN 2014C are likely related to the mass loss processes in the final stages of their progenitor system.

In brief, although the nebular spectra of these SNe presented in Figure \ref{nebu87A} share many similarities with that of SN 2014C, there are also notable differences. These discrepancies may be related to the transparency, composition, and structure of CSM, or the properties of explosion productions. Ultimately, they can be attributed to the different envelope-stripping processes of the progenitor before the explosion.

\begin{figure*}
\centering
\includegraphics[width=13cm,angle=0]{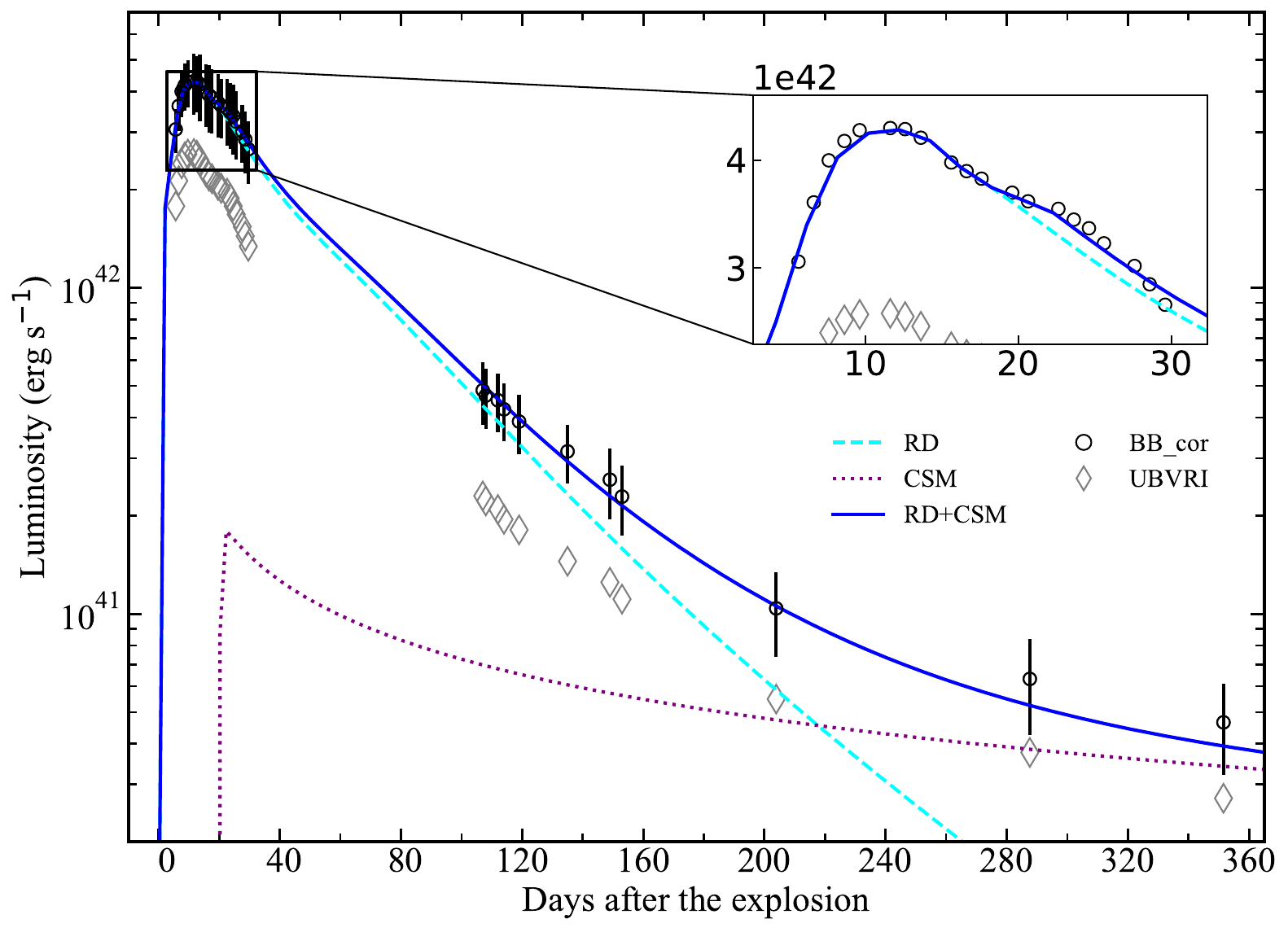}
\caption{Bolometric light curve of SN 2014C derived from the black body fit based on the $UBVRI$-band photometry. The integrated flux of $UBVRI$-band is also plotted. The dotted, dashed, and solid lines represent the radioactive decay (RD), CSM, and RD+CSM models, respectively.}
\label{LC2fit}
\end{figure*}

\begin{figure}
\centering
\includegraphics[width=8cm,angle=0]{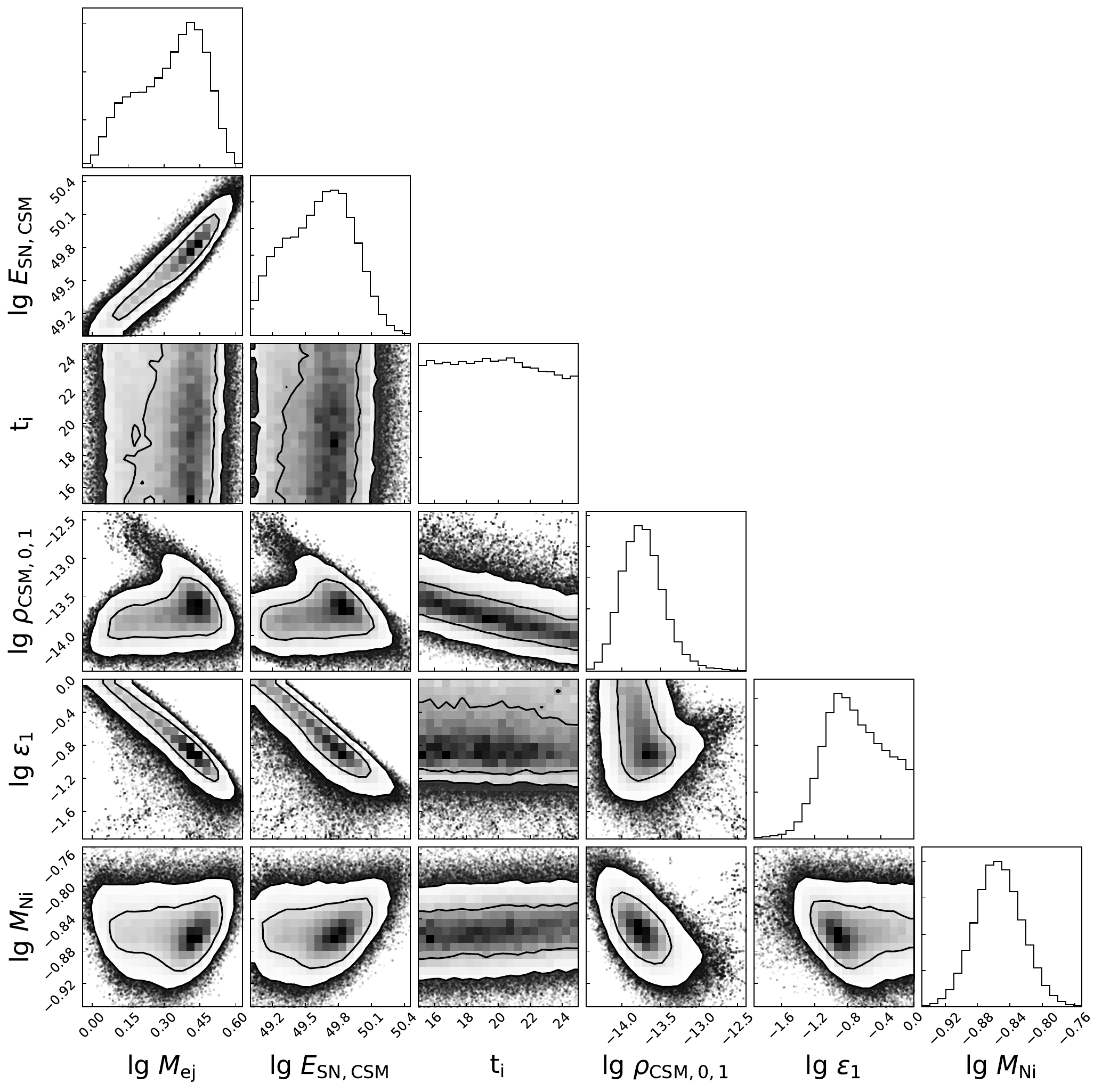}
\caption{Joint confidence level contours of the CSIRD parameters inferred from the MCMC-based fitting.}
\label{mcmc}
\end{figure}

\section{Bolometric light curve and modeling}
\label{bolo}

Figure \ref{bolos} presents the bolometric light curve of SN 2014C derived from the black body fitting of $UBVRI$-band photometry. The peak bolometric luminosity of SN 2014C, $L_{\rm max}\approx4.3\times10^{42}\rm erg\,s^{-1}$, is higher than typical SNe Ib values (e.g., $\sim 1.5\times10^{42}\rm erg\,s^{-1}$; \citealp{2020A&A...642A.106D}), with a rise time of $\sim$ 11.6 d, shorter than the usual $\sim$ 20 d for SNe Ib. Even within the SESNe family, the peak luminosity of SN 2014C remains on the high side. This figure compares it with a representative group of SESNe, showing that the peak luminosity of SN 2014C is comparable to that of SN Ib 1999ex and only lower than that of SN Ib/c 2005bf. The exceptional luminosity of SN 2005bf could be attributed to the energy supplied by a magnetar resulting from the explosion, given this SN may be associated with a gamma-ray burst \citep{2006ApJ...641.1039F}.

The bolometric luminosity of SN 2008D, when scaled by a factor of 2.2, aligns closely with that of SN 2014C during $15\,\lesssim\tau\,\lesssim\,30$ d. For $\tau\,\gtrsim\,30$ d, the luminosity evolution of SN 2014C may initially follow a trend similar to that of SN 2008D, but subsequently decreases at a slower rate. This partially fills the data gap for SN 2014C between $30\,\lesssim\,\tau\lesssim\,100$ d. Considering the uncertainty in the explosion time of SN 2014C, if we assume it reached its bolometric peak around 20 days, similar to SN 2008D, this could explain the similarity in their spectral features, particularly the velocity evolution of \HeI\,$\lambda$\,5876. However, even with this amplification factor, the luminosity of SN 2014C remains significantly higher than that of SN 2008D both before the peak and during the tail phase.

Following the radioactive decay (RD) model (e.g., Arnett law; \citealp{1982ApJ...253..785A,2005A&A...431..423S}), as seen in Eq. \ref{eqNi}, the mass of $^{56}$Ni produced during the explosion is $\rm M(^{56}Ni) = 0.14\,\pm\,0.03$\,\Msun, 
\begin{equation}
M_{\rm Ni}=\frac{{L_{\rm max}}}{10^{43} \rm erg\ s^{-1}} \times (6.45 e^{-{t_r \over \tau_{\rm Ni}}}+1.45 e^{-{t_r \over \tau_{\rm Co}}})^{-1},
\label{eqNi}
\end{equation}
where, $\tau_{\rm Ni} = 8.8$ d and $\tau_{\rm Co} = 111.3$ d are the decay time of \Nifs\ and $^{56}$Co.  This result aligns with the estimate by \cite{2017ApJ...835..140M}, who calculated 0.15 \Msun, accounting for their 0.2 mag overestimation in peak luminosity.

As shown in Figure \ref{LC2fit}, the bolometric light curve of SN 2014C at $\tau\lesssim\,15$ days can be well fitted by the RD model with $\rm M(^{56}Ni) = 0.14 $ \Msun. This suggests that SN 2014C could be mainly powered by radioactive decay during the early phase. The fast rise could be attributed to a small progenitor radius or a relatively high ratio of explosion energy to ejected mass. However, SN 2014C exhibits flux excess at $\tau\gtrsim\,20$ compared to this model, indicating the presence of an additional energy source apart from the radioactive decay of \Nifs. 

The flat and blue color curve seen in Figure \ref{color_curves} at $\tau\gtrsim\,20$, along with the absence of significant X-ray emissions during the first 20 days \citep{2017ApJ...835..140M}, supports the hypothesis of delayed interaction between SN and surrounding CSM. Previous observational evidence suggests that the interaction of SN 2014C began 45 d post-explosion and continued for more than 2000 d \citep{2016ApJ...833..231T, 2017ApJ...835..140M, 2015ApJ...815..120M, 2017MNRAS.466.3648A,2022ApJ...939..105B, 2022ApJ...930...57T}. Therefore, a hybrid model incorporating both radioactive decay and ejecta-CSM interaction (CSIRD) is adopted to explain the light curve of SN 2014C.

Assume that the CSM has a stellar-wind density profile (i.e. $\rho\propto r^{-2}$) and that the density of SN ejecta is uniform for the inner ejecta and follows $\rho\propto r^{-n}$ (here $n=7$) for the outer ejecta with the dimensionless transition radius of $x_0=0.3$. Here we develop the popular CSMRD model \citep{2012ApJ...746..121C, 2013ApJ...773...76C} by introducing a gamma-ray leakage factor of $A\approx  3n (n-3)x_0\kappa_\mathrm{Ni}M_\mathrm{ej}/[4 \pi (n-1)(n-3+3x_0^3) v_\mathrm{ej}^2 ]$ ($x_0^n<<1$). The modified CSMRD-powered luminosity can be rewritten as

\begin{equation}
\label{e:C12}
\begin{split}
L_\mathrm{SN}=&\frac{1}{t_0}\exp^{-t'/t_0}\int^{t}_0 \exp^{t'/t_0} P_\mathrm{CSI,inp} \mathrm{d} t' \\&+\frac{1}{t'_0}\exp^{-t'/t'_0}\int^{t}_0 \exp^{t'/t'_0} P_\mathrm{RD,inp} (1-\exp^{-At'^{-2}})\mathrm{d} t',
\end{split}
\end{equation}

, where the $t_0$ and $t'_0$ are the diffusion timescales through the total CSM mass ($M_\mathrm{CSM}$) and the CSM+SN mass ($M_\mathrm{CSM}+M_\mathrm{ej}$), $P_\mathrm{CSI,inp}$ and $P_\mathrm{Ni,inp}$ are the input powers from ejecta-CSM interaction \citep{2012ApJ...746..121C} and radioactive decay of $^{56}Ni+^{56}Co$ \citep{2008MNRAS.383.1485V}, and $\kappa_{\gamma}=0.027$~cm$^2$~g$^{-1}$ is the opacity to the gamma-ray photons from $^{56}$Ni and $^{56}$Co decay (e.g., \citealp{1997A&A...328..203C, 2000ApJ...545..407M, 2003ApJ...593..931M, 2016ApJ...817..132D}). The detection of interaction signals starting 20 days post-explosion indicates that the SN ejecta caught up with the H-rich CSM at that moment. Hence, we set the prior on the initial time of interaction as $t_\mathrm{i}=20\pm5$ days. The CSM mass is set to be 1 $M_\odot$, as inferred from X-ray and radio observations \citep{2017ApJ...835..140M,2015ApJ...815..120M}. Factors affecting the final fitting results, such as the structure of CSM and the efficiency of converting kinetic energy into radiation during the SN-CSM interaction, were not fully accounted for in this study. We used a spherically symmetric CSM structure rather than more complex configurations, such as the torus-shaped CSM proposed by \cite{2022ApJ...930...57T}. This simplification may lead to an overestimation of CSM mass \citep{2022ApJ...939..105B}. However, \cite{2024arXiv241017699O} utilized 3D hydrodynamic modeling to determine a CSM mass of 2.5 M$_\odot$ for SN 2014C, which exceeds the values reported in most recent research. Considering the uncertainties in CSM mass estimation, our intermediate value is deemed appropriate.

Based on a Markov chain Monte Carlo (MCMC) sampling algorithm \citep{2013PASP..125..306F}, the fitting with the CSIRD model suggests that the SN ejecta has a total mass of $(2.22_{-0.77}^{+0.69})\, M_\odot$, a $^{56}Ni$ mass of $(0.14_{-0.01}^{+0.01})\, M_\odot$, and more details are presented in Figure \ref{mcmc}. 
The CSM has an inner-radius density of $(1.81_{-0.78}^{+1.57})\times10^{-14}$ g cm$^{-3}$, and the initial time of interaction is $t_\mathrm{i} = 19.9_{-3.25}^{+3.34}$ days. The efficiency of interaction energy converted to radiation is $0.2_{-0.1}^{+0.3}$. If the CSM results from a mass ejection of the progenitor star before the explosion, the mass-loss rate can be estimated as $0.2/(v_\mathrm{CSM}/1000$ km s$^{-1}$) $M_\odot$ yr$^{-1}$, where $v_\mathrm{CSM}$ is the velocity of the CSM. Wolf-Rayet stars or red supergiant stars have difficulty in producing such intense ejections, while giant eruptions from a luminous blue variable star (LBV) can be responsible for such violent pre-SN mass ejection of $10^{-2}-10\,M_\odot$ yr$^{-1}$ (see \citealp{2014ARA&A..52..487S, 2019MNRAS.488.3783B} and references therein). Given the complex CSM revealed by our observations, and the progenitor observation \citep{2020MNRAS.497.5118S}, SN 2014C could originate from the explosion of an LBV in a binary system.

\section{Summary}
\label{sum}
This work presents high-cadence spectral and photometric observations of SN 2014C during the first month after its explosion. We achieved precise photometric results using the template subtraction method, while the analysis of over ten early spectra revealed detailed insights into ejecta evolution. These observations, along with related calculations and model analyses, improve the understanding of the nature of this metamorphic SN.

The light curve of SN 2014C exhibited a faster rise time ($\sim$\,11.5 d) than typical SNe Ib/c ($\gtrsim 16$ d) and reached a relatively higher peak luminosity ($L_{\rm max}\approx4.3\times10^{42}\rm erg s^{-1}$). The rapid rising luminosity could indicate a high ratio of explosion energy to ejecta mass. Based on the peak luminosity, we estimate that this SN synthesized 0.14 \Msun\,of \Nifs\,during the explosion. However, the energy released from the radioactive decay of this nickel is still insufficient to fully account for the light curve at $\tau\,\gtrsim\,20$\,d. To address this, we introduced an SN-CSM interaction model to provide additional energy.
Although this combined model fits the observed data reasonably well, it is important to acknowledge that it represents only one possible explanation. The SN-CSM interaction could have commenced earlier and intensified rapidly, leading to the swift rise in the light curve and influencing the peak luminosity. Consequently, the nickel mass derived from the peak bolometric luminosity may represent an upper limit, suggesting that the amount of nickel synthesized could be lower.

The decelerating and fading H$\alpha$ absorption, observed at a velocity higher than those of the photospheric components like Ca and He at $t \lesssim 16$ d, indicates that the hydrogen envelope of the progenitor was not completely stripped before the core collapse.

Through the earliest interaction signal detected in the photometry, the presence of CSM begins at a distance of $4 \times\,10^{14}$ cm from the SN. However, factors such as density or optical depth prevented the observation of significant SN-CSM interaction signals in the spectrum during this phase.
The emergence of intermediate H$\alpha$ at $t \approx$ 98 d suggests a denser CSM located at $\sim\,10^{16}$ cm. These observations imply variations in the CSM, as reported by \cite{2017ApJ...835..140M}. They found that the CSM had a low density at $R\lesssim2\times10^{16}$ cm and the dense H-rich material region at $R\sim5.5\times10^{16}$ cm.
Furthermore, the very long baseline interferometry (VLBI) observations have revealed a circular thin spherical shell as the structure of the H-rich CSM \citep{2018MNRAS.475.1756B,2021MNRAS.502.1694B,2022ApJ...930...57T}, and the CSM contains a mixture of carbonaceous and silicate and extends to at least 1.4$\times10^{17}$cm \citep{2019ApJ...887...75T}.

Given the presence of a small amount of hydrogen in the envelope and the nearby H-rich CSM with varying densities, it is evident that SN 2014C originated in a highly complex H-rich environment. This complex environment may be responsible for translating characteristics of SN 2014C among those of Ib, IIb, and IIn.
The transitional behavior of SN 2014C suggests that massive stars can explode at any phase during envelope stripping, and the state of stripping and the surrounding environment at the time of explosion contribute to the rich diversity observed in SESNe.\\


We thank the anonymous referee for providing a comprehensive report clarifying the paper and sparking fresh, qualitative insights. This work is supported by the National Key R\&D Program of China with No. 2021YFA1600404, the National Natural Science Foundation of China (12173082, 12333008), the science research grants from the China Manned Space Project with No. CMS-CSST-2021-A12, the Yunnan Fundamental Research Projects (grants 202401BC070007 and 202201AT070069), the Top-notch Young Talents Program of Yunnan Province, the Light of West China Program provided by the Chinese Academy of Sciences, and the International Centre of Supernovae, Yunnan Key Laboratory (No. 202302AN360001). EP is acknowledges financial support from INAF.

We acknowledge the support of the staff of the LJT, Asiago 1.82 m telescope, and NOT. Funding for the LJT has been provided by the CAS and the People's Government of Yunnan Province. The LJT is jointly operated and administrated by YNAO and the Center for Astronomical Mega-Science, CAS. 

%

\vspace{5mm}
\facilities{ LJT (+YFOSC), $Swift$-UVOT, Asiago 1.82 m Copernico telescope (+AFOSC), NOT (+ALFOSC)} 


\software{PyRAF \citep{2012ascl.soft07011S}, NumPy \citep{2020Natur.585..357H}, Matplotlib \citep{2007CSE.....9...90H}, Astropy \citep{2013A&A...558A..33A,2018AJ....156..123A,2022ApJ...935..167A}}


\bibliography{ms}
\bibliographystyle{aasjournal}


\begin{appendix}

\section{Photometric and spectroscopic data}
Table \ref{Tab:Photo_stand} enumerates the standard $UBVRI$ magnitudes for twelve local reference stars calibrated against the Landolt luminosity stars. Utilizing these reference stars, the instrumental magnitudes captured by LJT for SN 2014C were transformed into standard $UBVRI$ magnitudes, as detailed in Table \ref{Tab:Pho_Ground2}. The photometric outcomes from $Swift$-UVOT are summarized in Table \ref{Tab:Swiftpho}. The log of spectroscopic observations for SN 2014C, conducted in 2014, is outlined in Table \ref{Tab:Spec_log}. The photometric parameters of the sample presented in Figures \ref{light_curves} and \ref{color_curves} are collected in Table \ref{Tab:compared sam}. Lastly, the reference list for Figure \ref{comp} is provided in Table \ref{Tab: spec}.

\setcounter{table}{0} 
\renewcommand{\thetable}{\Alph{section}\arabic{table}}

\begin{table*}[!th]
\scriptsize
\centering
\caption{Photometry of the comparison stars in the field of SN 2014C$^a$}
\begin{tabular}{cccccccc}
\hline\hline
Star & R.A. & Dec. & $U$ (mag) & $B$ (mag) & $V$ (mag) & $R$ (mag) & $I$ (mag)  \\
\hline
1	&	22:37:12.236	&	+34:28:49.74	&	16.51(0.02)	&	16.41(0.01)	&	15.72(0.01)	&	15.34(0.01)	&	14.93(0.01)	\\
2	&	22:37:13.729	&	+34:28:27.68	&	17.64(0.03)	&	17.13(0.01)	&	16.25(0.01)	&	15.77(0.01)	&	15.30(0.01)	\\
3	&	22:37:11.107	&	+34:27:29.33	&	17.48(0.01)	&	17.31(0.01)	&	16.61(0.01)	&	16.24(0.01)	&	15.83(0.01)	\\
4	&	22:36:53.994	&	+34:27:19.53	&	16.57(0.02)	&	16.42(0.02)	&	15.73(0.01)	&	15.37(0.01)	&	14.98(0.01)	\\
5	&	22:37:20.518	&	+34:25:15.85	&	17.10(0.02)	&	16.67(0.01)	&	15.82(0.01)	&	15.35(0.01)	&	14.90(0.01)	\\
6	&	22:37:16.932	&	+34:24:28.56	&	18.12(0.03)	&	17.51(0.01)	&	16.62(0.01)	&	16.16(0.01)	&	15.71(0.01)	\\
7	&	22:37:22.365	&	+34:24:19.10	&	17.55(0.02)	&	16.68(0.01)	&	15.68(0.01)	&	15.12(0.01)	&	14.58(0.01)	\\
8	&	22:37:13.519	&	+34:23:24.97	&	16.76(0.03)	&	16.67(0.02)	&	15.96(0.01)	&	15.57(0.01)	&	15.16(0.01)	\\
9	&	22:37:21.993	&	+34:23:21.52	&	18.58(0.02)	&	17.75(0.02)	&	16.70(0.01)	&	16.07(0.01)	&	15.42(0.01)	\\
10	&	22:37:14.596	&	+34:23:07.57	&	16.44(0.03)	&	16.47(0.01)	&	15.89(0.01)	&	15.56(0.01)	&	15.20(0.01)	\\
11	&	22:37:18.447	&	+34:22:48.35	&	17.31(0.01)	&	16.80(0.01)	&	15.93(0.01)	&	15.48(0.01)	&	15.03(0.01)	\\
12	&	22:36:58.213	&	+34:22:05.50	&	17.45(0.01)	&	17.01(0.02)	&	16.12(0.01)	&	15.67(0.01)	&	15.22(0.01)	\\
\hline
\hline
\end{tabular}
\\
$^a$See Figure. \ref{<img>} for the finder chart of these reference stars. $UBVRI$ bands in the Vega magnitude system. Uncertainties (in parentheses) are $1\sigma$.

\label{Tab:Photo_stand}
\end{table*}

\begin{table*}[!th]
\scriptsize
\centering
\caption{LJT $UBVRI$-bands photometry of SN 2014C$^a$ in 2014}.
\begin{tabular}{cccccccc}
\hline\hline
Date (UT)& MJD & Epoch (d)$^b$ & $U$ (mag) & $B$ (mag) & $V$ (mag) & $R$ (mag)& $I$ (mag)\\
\hline
Jan. 07	&	56664.53	&	-5.44	&	16.72(0.02)	&	16.44(0.01)	&	15.44(0.01)	&	14.92(0.01)	&	14.45(0.01)	\\
Jan. 08	&	56665.53	&	-4.44	&	16.59(0.03)	&	16.24(0.01)	&	15.24(0.01)	&	14.71(0.01)	&	14.26(0.01)	\\
Jan. 08$^c$&56665.81    &   -4.16   &   ...         &   16.18(0.02) &   15.13(0.03)&   14.65(0.02) &   ...         \\
Jan. 09	&	56666.51	&	-3.46	&	16.56(0.03)	&	16.10(0.01)	&	15.10(0.01)	&	14.58(0.01)	&	14.08(0.01)	\\
Jan. 10	&	56667.53	&	-2.44	&	16.59(0.03)	&	16.08(0.01)	&	15.03(0.01)	&	14.48(0.01)	&	13.98(0.01)	\\
Jan. 11	&	56668.53	&	-1.44	&	16.63(0.03)	&	16.09(0.01)	&	14.99(0.01)	&	14.42(0.01)	&	13.90(0.01)	\\
Jan. 13	&	56670.53	&	0.56	&	16.94(0.04)	&	16.13(0.01)	&	14.94(0.01)	&	14.33(0.01)	&	13.84(0.04)	\\
Jan. 14	&	56671.50	&	1.53	&	16.92(0.03)	&	16.17(0.01)	&	14.95(0.01)	&	14.35(0.01)	&	13.80(0.01)	\\
Jan. 15	&	56672.53	&	2.56	&	17.16(0.07)	&	16.21(0.01)	&	14.98(0.01)	&	14.36(0.01)	&	13.79(0.01)	\\
Jan. 17	&	56674.52	&	4.55	&	17.26(0.03)	&	16.37(0.01)	&	15.06(0.01)	&	14.42(0.01)	&	13.80(0.01)	\\
Jan. 18	&	56675.52	&	5.55	&	17.38(0.02)	&	16.41(0.01)	&	15.09(0.01)	&	14.43(0.01)	&	13.82(0.01)	\\
Jan. 19	&	56676.50	&	6.53	&	17.36(0.03)	&	16.43(0.03)	&	15.12(0.01)	&	14.45(0.01)	&	13.83(0.01)	\\
Jan. 21	&	56678.50	&	8.53	&	17.43(0.05)	&	16.49(0.01)	&	15.16(0.01)	&	14.52(0.01)	&	13.84(0.01)	\\
Jan. 22	&	56679.54	&	9.57	&	17.41(0.03)	&	16.53(0.01)	&	...			&	...			&	...			\\
Jan. 24	&	56681.52	&	11.55	&	17.57(0.03)	&	16.57(0.01)	&	15.24(0.01)	&	14.54(0.01)	&	13.88(0.01)	\\
Jan. 25	&	56682.53	&	12.56	&	17.62(0.03)	&	16.63(0.02)	&	15.29(0.01)	&	14.56(0.01)	&	13.90(0.01)	\\
Jan. 26	&	56683.52	&	13.55	&	17.79(0.03)	&	16.72(0.03)	&	15.35(0.01)	&	14.59(0.01)	&	13.91(0.01)	\\
Jan. 27	&	56684.50	&	14.53	&	17.87(0.08)	&	16.81(0.02)	&	15.43(0.01)	&	14.63(0.01)	&	13.94(0.01)	\\
Jan. 29	&	56686.50	&	16.53	&	...			&	16.95(0.03)	&	15.55(0.01)	&	14.71(0.01)	&	13.96(0.01)	\\
Jan. 30	&	56687.49	&	17.52	&	18.04(0.13)	&	17.01(0.03)	&	...			&	...			&	14.02(0.01	\\
Jan. 31	&	56688.50	&	18.53	&	18.12(0.09) &	17.06(0.06)	&	15.72(0.01)	&	14.86(0.01)	&	14.10(0.01)	\\
Apr. 18	&	56765.91	&	95.94	&	20.23(0.19)	&	19.18(0.09)	&	17.60(0.03)	&	16.68(0.03)	&	15.94(0.03)	\\
Apr. 19	&	56766.91	&	96.94	&	20.21(0.17)	&	19.21(0.12)	&	...			&	...			&	...			\\
Apr. 23	&	56770.90	&	100.93	&	...			&	...			&	17.73(0.03)	&	16.77(0.03)	&	15.99(0.03)	\\
Apr. 25	&	56772.92	&	102.95	&	...			&	...			&	17.82(0.04)	&	...	        &	16.06(0.02)	\\
Apr. 30	&	56777.90	&	107.93	&	...			&	19.48(0.08)	&	17.88(0.04)	&	16.93(0.03)	&	16.16(0.04)	\\
May 16	&	56793.90	&	123.93	&	20.79(0.12)	&	19.79(0.12)	&	18.14(0.03)	&	17.16(0.04)	&	16.38(0.03)	\\
May 30	&	56807.88	&	137.91	&	...			&	...			&	18.29(0.04)	&	17.26(0.03)	&	16.68(0.03)	\\
June 03	&	56811.89	&	141.92	&	...			&	...			&	18.42(0.05)	&	17.39(0.03)	&	16.81(0.03)	\\
July 24	&	56862.80	&	192.83	&	21.88(0.24)	&	20.84(0.15)	&	19.18(0.08)	&	18.08(0.04)	&	17.78(0.11)	\\
Oct. 16	&	56946.54	&	276.57	&	...			&	...			&	19.39(0.08)	&	18.55(0.06)	&	18.48(0.09)	\\
Nov. 11$^d$& 56972.99   &   303.02  &   ...         &   ...         &   19.60(0.04) &   18.63(0.04) &   ...         \\
Dec. 19	&	57010.48	&	340.51	&	...			&	...			&	19.88(0.26)	&	18.85(0.09)	&	18.78(0.12)	\\
\hline
\hline
\end{tabular}
\\
$^a${Uncertainties (in parentheses) are $1\sigma$.}\\
$^b${The epoch is relative to the $V$-band maximum date, MJD = 56669.97.}\\
$^c${Photometry taken by Asiago 1.82 m telescope +AFOSC.}\\
$^d${Photometry taken by NOT+ALFOSC.}

\label{Tab:Pho_Ground2}
\end{table*}

\begin{table*}
\scriptsize
\centering
\caption{{\it Swift} UVOT photometry of SN 2014C$^a$ in 2014}
\begin{tabular}{lcccccc}
\hline\hline
Date (UT) & MJD      & Epoch (d)$^b$ & $uvw1$(mag) & $u$(mag)      & $b$(mag)      & $v$(mag)\\
\hline
Jan. 06	&	56663.25	&	-6.72	&	...			&	17.07(0.16)	&	17.01(0.14)	&	15.79(0.12)	\\
Jan. 07	&	56664.21	&	-5.76	&	18.62(0.26)	&	17.22(0.16)	&	16.73(0.11)	&	15.59(0.10)	\\
Jan. 09	&	56666.49	&	-3.48	&	18.82(0.21)	&	16.65(0.11)	&	16.11(0.08)	&	15.08(0.07)	\\
Jan. 09	&	56666.52	&	-3.45	&	...			&	...			&	16.06(0.07)	&	15.10(0.07)	\\
Jan. 11	&	56668.45	&	-1.52	&	...			&	16.85(0.12)	&	16.09(0.08)	&	14.96(0.07)	\\
Jan. 11	&	56668.54	&	-1.43	&	18.90(0.23)	&	16.84(0.12)	&	16.04(0.07)	&	14.93(0.08)	\\
Jan. 13	&	56670.44	&	0.47	&	19.00(0.25)	&	17.03(0.13)	&	16.04(0.07)	&	14.93(0.06)	\\
Jan. 14	&	56671.46	&	1.49	&	...			&	...			&	16.06(0.07)	&	14.97(0.06)	\\
Jan. 15	&	56672.43	&	2.46	&	...			&	17.29(0.15)	&	16.16(0.08)	&	14.93(0.06)	\\
Jan. 15	&	56672.91	&	2.94	&	19.24(0.29)	&	16.93(0.12)	&	16.11(0.08)	&	15.05(0.07)	\\
Jan. 17	&	56674.35	&	4.38	&	...			&	17.33(0.16)	&	16.26(0.08)	&	15.06(0.07)	\\
Jan. 18	&	56675.10	&	5.13	&	19.75(0.44)	&	17.40(0.16)	&	16.35(0.08)	&	15.14(0.07)	\\
Jan. 19	&	56676.24	&	6.27	&	...			&	17.57(0.18)	&	16.48(0.09)	&	15.10(0.07)	\\
Jan. 19	&	56676.50	&	6.53	&	19.49(0.35)	&	17.37(0.16)	&	16.39(0.08)	&	...	\\
\hline
\hline
\end{tabular}

$^a${Uncertainties (in parentheses) are $1\sigma$.}\\
$^b${The epoch is relative to the $V$-band maximum date, MJD = 56669.97.}
\label{Tab:Swiftpho}
\end{table*}

\begin{table*}
\caption{Journal of spectroscopic observations of SN 2014C in 2014}
\scriptsize
\centering

\begin{tabular}{ccccccccc}
\hline\hline
Date& MJD & Epoch$^a$ & Range & Disp.  &Slit Width  & Exp. Time &airmass & Telescope\\
(UT) &  &(d) & (\AA) & (\AA\,pix$^{-1}$) &(pix) & (s) &  & +Instrument \\
\hline
Jan. 05	&	56662.53	&	-7.44	&	3800-8750	&	2.85	&	6.36	&	1800	&	1.54	&	LJT+YFOSC(G3)	\\
Jan. 06	&	56663.51	&	-6.46	&	3900-8750	&	2.85	&	6.36	&	2700	&	1.36	&	LJT+YFOSC(G3)	\\
Jan. 07	&	56664.49	&	-5.48	&	3550-8750	&	2.85	&	6.36	&	2700	&	1.26	&	LJT+YFOSC(G3)	\\
Jan. 08	&	56665.49	&	-4.48	&	3600-8750	&	2.85	&	6.36	&	2700	&	1.27	&	LJT+YFOSC(G3)	\\
Jan. 08	&	56666.28	&	-3.69	&	3370-8150 	&	4.84	&	3.25	&	1800	&	1.48	&	1.82m+AFOSC(G4)	\\
Jan. 09	&	56666.51	&	-3.46	&	3600-8750	&	2.85	&	6.36	&	2700	&	1.45	&	LJT+YFOSC(G3)	\\
Jan. 10	&	56667.48	&	-2.49	&	5240-9200	&	1.50	&	6.36	&	3600	&	1.28	&	LJT+YFOSC(G8)	\\
Jan. 12	&	56669.50	&	-0.47	&	4100-8750	&	2.85	&	6.36	&	2700	&	1.40	&	LJT+YFOSC(G3)	\\
Jan. 13	&	56670.49	&	0.52	&	5830-8120	&	0.84	&	3.53	&	2700	&	1.35	&	LJT+YFOSC(E13)	\\
Jan. 15	&	56672.50	&	2.53	&	3900-8750	&	2.85	&	6.36	&	1800	&	1.48	&	LJT+YFOSC(G3)	\\
Jan. 16	&	56673.54	&	3.57	&	4100-8750	&	2.85	&	6.36	&	1800	&	1.99	&	LJT+YFOSC(G3)	\\
Jan. 18	&	56675.49	&	5.52	&	3500-8750	&	2.85	&	6.36	&	1800	&	1.45	&	LJT+YFOSC(G3)	\\
Jan. 19	&	56676.51	&	6.54	&	3490-8750	&	2.85	&	6.36	&	1800	&	1.68	&	LJT+YFOSC(G3)	\\
Jan. 22	&	56679.51	&	9.54	&	3500-8750	&	2.85	&	6.36	&	2100	&	1.75	&	LJT+YFOSC(G3)	\\
Jan. 25	&	56682.49	&	12.52	&	3750-8750	&	2.85	&	6.36	&	2400	&	1.69	&	LJT+YFOSC(G3)	\\
Jan. 28	&	56685.50	&	15.53	&	3800-8750	&	2.85	&	6.36	&	2333	&	1.91	&	LJT+YFOSC(G3)	\\
Apr. 20	&	56767.89	&	97.92	&	5180-9100	&	3.66	&	6.36	&	1500	&	1.92	&	LJT+YFOSC(G5)	\\
Apr. 21	&	56768.89	&	98.92	&	4800-9140	&	7.72	&	6.36	&	1500	&	1.90	&	LJT+YFOSC(G10)	\\
May 16	&	56793.88	&	123.91	&	4080-9140	&	7.72	&	6.36	&	1500	&	1.34	&	LJT+YFOSC(G10)	\\
Jul. 24	&	56862.69	&	192.72	&	4300-9140	&	7.72	&	6.36	&	3530	&	1.32	&	LJT+YFOSC(G10)	\\
Nov. 11	&	56973.46	&	303.49	&	3000-9130   &	2.99	&	3.10	&	3600	&	1.15	&	NOT+ALFOSC(G4)	\\
\hline  
\hline
\end{tabular}

$^a${The epoch is relative to the $V$-band maximum date, MJD = 56669.97}
\label{Tab:Spec_log}
\end{table*}


\begin{table*}
\centering
\scriptsize
\caption{The detailed information of compared stars in Figure \ref{light_curves} and Figure \ref{color_curves}}
\begin{tabular}{ l c c c c c c c l}
\hline\hline
Star         &   DM   & $E(B-V)$ &$t^R_{rise}$ & $t^R_{1/2}$ &$M_{max}^R$  &$\Delta m_{15}(R)$ & type  &   Reference        \\
             & (mag)  & (mag)    & (d)         & (d)         & (mag)       & (mag)             &       &  \\
\hline
iPTF13bvn	   & 31.76  & 0.07     & 19.36       & 12.00     & -17.00	   &	1.09            &  Ib    &  1. 2. 3. 4. 5.    \\ 
SN 2005bf$^a$  & 34.50  & 0.05     & 42.47       & 18.68     & -18.18	   &	0.37            &  Ib/c  &  6. 7. 8. 9.  \\
SN 2007Y$^a$   & 31.43  & 0.11     & 20.14       & 10.39     & -16.43      &	0.81	        & Ib/IIb &  5. 10. \\ 
SN 2008D       & 32.46  & 0.60     & 19.60       & 13.18     & -17.14      &	0.15            &	Ib	 &	5. 6. 11. 12. \\ 
SN 2009jf	   & 32.65  & 0.16     & 21.44       & 14.08     & -18.12      &	0.31            &	Ib	 &	6. 13. 14. \\ 
SN 2011fu      & 34.36  & 0.22     & 26.40       & 14.65     & -18.35      &	0.51	        &	IIb	 &	15. \\ 
SN 2005hg      & 34.87  & 0.09     & 16.97       & 12.18     & -18.19      &	0.49	        &	Ib/c &	6. 16. 17 \\ 
SN 2008ax      & 29.92  & 0.30     & 21.80       & 10.85     & -17.57	   &	0.74	        &	IIb	 &	5. 18. 19. \\
SN 2019yvr$^a$ & 30.84  & 0.56     & 17.64       & 11.27     & -17.23      &	0.70            &	Ib	 &	20. \\
SN 2014C       & 30.89  & 0.70     & 12.13       & 7.28      & -18.22      &    0.33            &   Ib   &  This work \\
\hline
\hline
\end{tabular}
\\$^a${The samples use $r$ band magnitude in this table.}\\
1. \cite{2014AJ....148...68B} 2. \cite{2013ApJ...775L...7C} 3. \cite{2014A&A...565A.114F} 4. \cite{2016ApJ...825L..22F} 5. \cite{2014Ap&SS.354...89B}
6. \cite{2014ApJS..213...19B}  7. \cite{2006ApJ...641.1039F} 8. \cite{2007ApJ...666.1069M} 9. \cite{2018A&A...609A.134S}
10. \cite{2009ApJ...696..713S} 11. \cite{2008Sci...321.1185M} 12. \cite{2009ApJ...702..226M} 
13. \cite{2011MNRAS.416.3138V} 14. \cite{2011MNRAS.413.2583S} 
15. \cite{2013MNRAS.431..308K} 16. \cite{2011ApJ...741...97D} 17. \cite{2014AJ....147...99M} 
18. \cite{2008MNRAS.389..955P} 19. \cite{2009PZ.....29....2T} 20. \cite{2024MNRAS.529L..33F}

\label{Tab:compared sam}
\end{table*}

\begin{table*}
\centering
\scriptsize
\caption{Reference of the SNe in Figure. \ref{comp}}
\begin{tabular}{ccccc}
\hline\hline
Star               &(a) pre-maximum     & (b) maximum              & (c) 7 d post maximum          & (d) 14 days post maximum\\
\hline
iPTF13bvn	       &\cite{2013ApJ...775L...7C}          &	   \cite{2013ApJ...775L...7C}          &              ...              &  \cite{2014MNRAS.445.1932S}         \\
SN 2004gq         &\cite{2014AJ....147...99M}       &	   \cite{2014AJ....147...99M}       &              ...              &  \cite{2014AJ....147...99M}             \\
SN 2005bf     	   &\cite{2014AJ....147...99M}       &	   \cite{2006ApJ...641.1039F}    & \cite{2014AJ....147...99M}                 &  ...            \\
SN 2007Y     	   &\cite{2009ApJ...696..713S}  &	   ...                 & \cite{2009ApJ...696..713S}            & \cite{2019MNRAS.482.1545S}             \\
SN 2008D     	   &\cite{2019MNRAS.482.1545S}     &	   \cite{2009AIPC.1111..627M}     & \cite{2019MNRAS.482.1545S, 2009ApJ...702..226M}   & \cite{2014AJ....147...99M}             \\
SN 1997X     	   &     ...            &	   \cite{2014AJ....147...99M}       & \cite{2014AJ....147...99M}                 & ...             \\
SN 1999ex     	   &     ...            &	   \cite{2002AJ....124..417H}        & Superfit$^a$                  &\cite{2002AJ....124..417H}             \\
SN 2007C     	   &     ...            &	   \cite{2019MNRAS.482.1545S}     & \cite{2014AJ....147...99M}                 & \cite{2019MNRAS.482.1545S}             \\

\hline
\hline
\end{tabular}

$^a$http://www.dahowell.com/superfit.html
\label{Tab: spec}
\end{table*}


\end{appendix}




\end{document}